\documentstyle[seceq,preprint]{jpsj}

\newcommand{\ket}[1]{\left| \mbox{$\displaystyle #1$} \right>}
\newcommand{\bracket}[1]{\left[ \mbox{$\displaystyle #1$} \right]}
\newcommand{\parenthesis}[1]{\left( \mbox{$\displaystyle #1$} \right)}
\newcommand{\diff}[2]{\frac{\partial #1}{\partial #2}}
\newcommand{\sech}{{\rm sech}}
\newcommand{\veck}{{\bf k}}
\newcommand{\vecr}{{\bf r}}
\newcommand{\e}{{\rm e}}
\renewcommand{\i}{{\rm i}}
\renewcommand{\Im}{{\rm Im}}
\renewcommand{\Re}{{\rm Re}}
\def\lsim{\mathop {\vtop {\ialign {##\crcr
$\hfil \displaystyle {<}\hfil $\crcr \noalign {\kern1pt \nointerlineskip }
$\,\,\sim$ \crcr \noalign {\kern1pt}}}}\limits}

\title
{
A Theory of Anisotropic Semiconductor of Heavy Fermions
}

\author
{
Hiroaki {\sc Ikeda} and Kazumasa {\sc Miyake}
}

\inst
{
Department of Material Physics, Faculty of Engineering Science,
 Osaka University, Toyonaka 560
}

\recdate
{
November 24, 1995
}

\abst
{
It is demonstrated that a {\veck}-dependence of the hybridization
matrix element between $f$- and conduction electrons can give rise to
an anisotropic hybridization gap of heavy fermions if the filling of
electrons corresponds to that of the band insulator. The most interesting
case occurs when the hybridization vanishes along some symmetry axis
of the crystal reflecting a particular symmetry of the crystal field.
The results of a model calculation are consistent with wide range of
anomalous properties observed in CeNiSn and its isostructural compounds,
the anisotropic semiconductor of heavy fermions. In particular, highly
sensitive effect of impurity scattering on the residual density of
states for zero energy excitation and the anisotropic temperature
dependence of the resistivity are well explained. It is also discussed
that a weak semimetallic behavior arises through the weak $\veck$-dependence
of the $f$-electron self-energy $\Sigma_{f}(\veck,0)$.
}

\kword
{heavy fermion, Kondo insulator, anisotropic hybridization gap,
crystal field, CeNiSn, CeRhSb}

\begin{document}
\sloppy
\maketitle

\newpage

\section{Introduction}

Heavy-fermion systems exhibit a variety of ground states. Among them
there is a class, the so-called  ``Kondo insulator'', which has a narrow
energy gap of the order ($\sim 10 K$) at low temperatures. This class of
compounds exhibits the Kondo effect at high temperatures and an insulating
behavior at low temperatures.~\cite{rf:Falicov,rf:Oomi} The mechanism of
the energy-gap formation at low temperatures has been discussed from
a variety of view points. These are classified roughly into two categories,
i.e., {\veck}-space and real-space approaches. In the former approach,
the origin of insulating behaviors is attributed to a hybridyzation gap
which is highly renormalized by the strong correlations among almost
localized $f$-electrons.~\cite{rf:Martin} A circumferential evidence
supporting this point of view is that all the compounds called the ``Kondo
 insulator'', except for TmSe, have even number of electrons in the unit
cell which is a necessary condition for the band insulator. In the latter
approach, on the other hand, it is attributed to the formation of local
bound state of one kind of another, such as local singlet due to the strong
Kondo effect, the Wigner crystal formation, and so on.~\cite{rf:Kasuya,rf:ohta}

A picture of the renormalized hybridization gap is based on the principle
of ``adiabatic continuation'' \cite{rf:Anderson} which was applied first
by Landau in the Fermi liquid theory~\cite{rf:Landau} and then has turned
out very useful so far in understanding the low energy properties of wide
range of strongly interacting systems as demonstrated by Yamada and Yosida
for the impurity Anderson model.~\cite{rf:Yamada1} While correctness of this
picture for the  ``Kondo insulator'' has been suggested by the Gutzwiller
approach for the periodic Anderson model for this
decade,~\cite{rf:Rice,rf:Shiba} it was recently shown more vividly on
the basis of the numerical renormalization group method~\cite{rf:Sakai},
the quantum Monte Carlo calculation~\cite{rf:Mutou} and the exact
diagonalization method~\cite{rf:Saso} with the help of the $d=\infty$
theory.
On the other hand, it is reported that a charge gap is different from a
spin gap in the $d=1$ theory on the basis of the exact diagonalization
method.~\cite{rf:Ueda} This result is against the picture mentioned above.
However, we believe that this may come from the particularity of
1-dimensionality.

Among the  ``Kondo insulators'', CeNiSn and its isostructural compounds have
attracted much attention for almost a decade, because they have exhibited,
at least for samples of early stage, behaviors of anisotropic semiconductor
with highly renormalized energy gap.~\cite{rf:Takabatake1,rf:Malik,rf:Yoshii}
An anisotropic semiconductor means that the energy gap vanishes in a direction
in the {\veck}-space and the density of states (DOS) shows a V-shaped like
structure at low energy region. This was first inferred from the measurements
of the longitudinal NMR relaxation rate
at low temperature~\cite{rf:Kyogaku,rf:Nakamura,rf:Ohama} and reinforced later
by those of the thermodynamic
properties~\cite{rf:Takabatake2,rf:Nakamoto,rf:Nishigori,rf:Suderow}
(the specific heat, the magnetic susceptibility etc.). In addition, the
inelastic neutron scatterings show the existence of anisotropic magnetic
excitations in the $\veck$-space.~\cite{rf:Mason,rf:Kadowaki} These compounds
are also found to be very sensitive to a small amount of impurities such that
the residual DOS at zero energy excitations increases drastically in roughly
proportional to the square root of impurity concentration.

In this paper we develope a theory of $\veck$-space approach to understand
the anomalous properties of anisotropic semiconductor of heavy fermions, such
as CeNiSn. We follow the formalism developed for the Fermi liquid theory of
heavy fermions on the basis of the periodic Anderson model,~\cite{rf:Yamada2}
while we apply it to the filling corresponding to the band insulator.
An essential point of our theory is that the hybridization matrix element can
happen to vanish along some symmetry axis (in the $\veck$-space) of the crystal
for a particular symmetry of the crystal field which is expected to realize
in CeNiSn. Then, the hybridization gap also vanishes along the same direction,
which can explain wide range of the anomalies observed in CeNiSn.

An idea that the node of the hybridization can possibly give rise to the
pseudo-gap structure of quasiparticles has been addressed by Kagan, Kikoin
and Prokof'ev,~\cite{rf:Kagan} although their theory seems not to have been
fully developed and includes some ambiguity. For instance, the condition for
the appearance of the node of the hybridization was neither specified, nor
the explicit form of the quasiparticle DOS was given. Hereafter we develop
the theory on a more solid ground of the formalism.

We develope the formalism of our theory in $\S${\ref{sec:Hamiltonian}} $\sim$
 $\S${\ref{sec:Impurity}}, and discuss about DOS in $\S${\ref{sec:DOS}} and
the effect of impurity scattering on DOS in $\S${\ref{sec:ImpurityDOS}}.
$\S${\ref{sec:Physical}} is devoted to the discussions of physical properties
and those validity: the specific heat ($\S${\ref{sec:SpecificHeat}}), the
longitudinal NMR relaxation rate ($\S${\ref{sec:NMR}}), the magnetic
properties ($\S${\ref{sec:Magnetization}} and $\S${\ref{sec:Susceptibility}}),
the neutron scattering intensity ($\S${\ref{sec:Neutron}}), and the anisotropic
temperature dependence of the resistivity ($\S${\ref{sec:Resistivity}}) are
discussed in detail on the model calculations. Futhermore, the effect of
pressure ($\S${\ref{sec:Pressure}}) and the lifetime of quasiparticles
($\S${\ref{sec:Lifetime}}) are briefly discussed.

\section{Theory}

\subsection{Hamiltonian}
\label{sec:Hamiltonian}

We start with the periodic Anderson model keeping it in mind that
$(4f)^1$ configuration is realized in Ce$^{3+}$ ion in those
compounds:~\cite{rf:Yamada3}
\begin{equation}
  H = H_{c} + H_{f} + H_{c-f},
\label{eqn:Hamiltonian}
\end{equation}
where $H_{c}$, $H_{f}$, and $H_{c-f}$ stands for the Hamiltonian of conduction
electrons, $f$-electrons, and the hybridization among them, respectively.
The first term in eq.(\ref{eqn:Hamiltonian}), $H_{c}$, is given by
\begin{equation}
  H_{c} = \sum_{\veck\sigma}\xi_{\veck}c^{\dag}_{\veck\sigma}c_{\veck\sigma},
\end{equation}
where $c^{\dag}_{\veck\sigma} (c_{\veck\sigma})$ creates (annihilates) a
conduction electron in a plane wave state labeled by wave vector $\veck$
and spin $\sigma(\pm)$. The plane wave state can be expanded around site
$i$ ($\vecr_{i}$) as follows:
\begin{equation}
\label{eqn:plane-wave}
  \ket{\veck\sigma}=\dfrac{1}{\sqrt{V}}\e^{\i\veck\cdot\vecr}\chi_{\sigma}
                   =\dfrac{4\pi}{\sqrt{V}}\e^{\i\veck\cdot\vecr_{i}}
		          \sum^{\infty}_{l=0}\i^{l}j_{l}(k|\vecr-\vecr_{i}|)
		          \sum^{l}_{m=-l}Y^{m*}_{l}(\Omega_{\veck})
		          Y^{m}_{l}(\Omega_{\vecr-\vecr_{i}})\chi_{\sigma},
\end{equation}
where $\chi_{\sigma}$ is the spin function, $j_{l}(kr)$ is the spherical
Bessel function, $Y^{m}_{l}$ is the spherical harmonics with the argument
of solid angle $\Omega_\vecr$ of the position vector {\vecr} or
$\Omega_{\veck}$ of the wave vector {\veck}, and V is the volume of the
crystal.

The second term in eq.(\ref{eqn:Hamiltonian}), $H_{f}$, is given by
\begin{equation}
\label{eqn:atomic}
H_{f} = \sum_{{\veck}M}E_{M}f^{\dag}_{{\veck}M}f_{{\veck}M}
  +\dfrac{1}{2}U\sum_{i,M \neq M'}f^{\dag}_{iM}f_{iM}f^{\dag}_{iM'}f_{iM'},
\end{equation}
where $f^{\dag}_{iM} (f_{iM})$ is the creation (annihilation) operator of
the $f$-electron on the orbital $M$ in the $4f$ shell at site $i$, and
$f^{\dag}_{\veck M} (f_{\veck M})$ is its Fourier transform, and $E_{M}$'s
denote the energy levels of the $4f$-electron which are split into $j$=7/2
and $j$=5/2 multiplets under the spin-orbit interaction and further separated
by the crystal-field interaction in general. The angular part of its eigen
function is expressed as
\begin{equation}
  \ket{M} = \sum_{\mu}b^{M}_{\mu}\sum_{m\sigma}a^{\mu}_{lm\sigma}
             Y^{m}_{l}(\Omega_{\vecr-\vecr_{i}})\chi_{\sigma},
\label{eqn:f-state}
\end{equation}
where $\mu$ is the $z$-component of the total angular momentum, $j$, and
$a^{\mu}_{lm\sigma}$ are the Clebsch-Gordan coefficients, and $b^{M}_{\mu}$
are coefficients specifying the crystal-field level. The last term in
eq.(\ref{eqn:atomic}) represents the Coulomb repulsion $U$ between
$f$-electrons in the states $\ket{M}$ and $\ket{M'}$. For simplicity,
we neglect $M$- and $M'$- dependence of $U$.

The last term in eq.(\ref{eqn:Hamiltonian}), $H_{c-f}$, describes the
hybridization between $f$- and conduction electrons:
\begin{equation}
  H_{c-f} = \sum_{\veck M\sigma}(V_{\veck M\sigma}
		c^{\dag}_{\veck\sigma}f_{\veck M}+h.c.),
\end{equation}
where $V_{\veck M\sigma}$ is the mixing matrix element which can be calculated
with the use of eqs.(\ref{eqn:plane-wave}) and (\ref{eqn:f-state}) as
\begin{equation}
  V_{\veck M\sigma}=\sqrt{4\pi}V_{kl}\sum_{\mu}b^{M}_{\mu}\sum_{m}
                       a^{\mu}_{lm\sigma}Y^{m}_{l}(\Omega_{\veck}).
\end{equation}
Here, $V_{kl}$ has the energy dependence of the mixing matrix, and is treated
as a parameter of our model.

\subsection{Hybridization and Green Function}
\label{sec:Hybridization}

Now we consider the case in which the crystal-field splitting is so large that
we can neglect the effects of excited crystal-field states in the relevant
low-temperature and low-energy phenomena. Then we are left with two conduction
bands ($\pm\sigma$) and two $f$-levels($\pm M$). In this case the Green
functions of conduction- and $f$-electrons are given
by~\cite{rf:Yamada2,rf:Yamada3}
\begin{subequations}
\label{eqn:Green}
\begin{equation}
  G^{c}_{\sigma}(\veck,\omega)
     = \dfrac{1}{\omega-\xi_{k}
      -\dfrac{V^{2}_{f}(\veck)}{\omega-E_{f}-\Sigma_{f}(\veck,\omega)}} ,
\label{eqn:Gc}
\end{equation}
\begin{equation}
  G^{f}_{M}(\veck,\omega)
    = \dfrac{1}{\omega-E_{f}-\Sigma_{f}(\veck,\omega)
     -\dfrac{V^{2}_{f}(\veck)}{\omega-\xi_{\veck}}},
\label{eqn:Gf}
\end{equation}
\end{subequations}
where $E_{f}$ is the lowest crystal-field level and $\Sigma_{f}(\veck,\omega)$
is the self-energy of $f$-electrons due to the Coulomb repulsion $U$. The
hybridization $V^{2}_{f}(\veck)$ can be regarded as independent of $\sigma$
after an appropriate linear combinations of $\ket{M}$ and $\bar{\ket{M}}$ have
been taken.
\begin{equation}
V^{2}_{f}(\veck)
 \equiv |V_{\veck M\sigma}|^{2}+|V_{\veck\bar{M}\sigma}|^{2}
   =    |V_{\veck M\sigma}|^{2}+|V_{\veck M\bar{\sigma}}|^{2}
\end{equation}
The {\veck}-dependence of $V_{f}(\veck)$ reflects the symmetry of the lowest
crystal-field level in general.~\cite{rf:Yamada3,rf:Hanzawa} For example, in
the case of CeNiSn and its isostructual compounds, in which the approximate
local symmetry of the crystal field is trigonal $D_{3d}$~\cite{rf:Symmetry},
the energy levels split into three doublets: $\ket{5/2,\pm 3/2},
 a\ket{5/2,\pm 1/2}+b\ket{5/2,\mp 5/2}$, and
 $-b\ket{5/2,\pm 1/2}+a\ket{5/2,\mp 5/2}$, with $a$ and $b$ being appropriate
constants satisfying $a^{2}+b^{2}=1$. Depending on the symmetry of the
crystal-field ground state, there occurs various angular dependence of the
hybridization $V^{2}_{f}(\veck)$. While the hybridization is finite at any
direction of $\hat{k} \equiv \veck/|\veck|$ in the state including
$\ket{5/2,\pm 1/2}$, such as
$\ket{\pm p} \equiv a\ket{5/2,\pm 1/2}+b\ket{5/2,\mp 5/2}$, it vanishes along
the quantization axis ($z$-axis) in
$\ket{\pm m} \equiv \ket{5/2,\pm 3/2}$:~\cite{rf:Hanzawa}
\begin{subequations}
\begin{equation}
V^{2}_{\pm p}(\veck)
= V^{2}\left[ a^{2}2(5\hat{k}_{z}^{4}-2\hat{k}_{z}^{2}+1)
  +b^{2}5(1-\hat{k}_{z}^{2})^{2}
-4\sqrt{10}ab(\hat{k}_{x}^{2}-3\hat{k}_{y}^{2})\hat{k}_{x}\hat{k}_{z}\right],
\end{equation}
\begin{equation}
\label{eqn:V2}
V^{2}_{\pm m}(\veck) = V^{2}(1-\hat{k}_{z}^{2})(1+15\hat{k}_{z}^{2}),
\end{equation}
\end{subequations}
where $V^{2}$ gives $|\veck|$-dependence of $V^{2}(\veck)$ and $z$-axis is
taken as parallel to the $a$-axis, the symmetry axis of these crystals.

\subsection{Effective Hamiltonian for Quasiparticles}

We are interested in the low temperature region, in which the physical
properties can be described by the renormalized quasiparticles near the Fermi
level after the many-body effect due to the on-site repulsion $U$ in
eq.(\ref{eqn:atomic}) has been taken into account. These quasiparticles are
described by the effective Hamiltonian
\begin{subequations}
\begin{equation}
  \tilde{H}_{eff}=\sum_{\veck\sigma}\xi_{\veck}
                    c^{\dag}_{\veck\sigma}c_{\veck\sigma}
		 +\sum_{\veck}\tilde{E}_{f}
		    \tilde{f}^{\dag}_{\veck}\tilde{f}_{\veck}
		 +\sum_{\veck\sigma}\sum_{M=\pm}(\tilde{V}_{\veck M\sigma}
		    c^{\dag}_{\veck\sigma}\tilde{f}_{\veck}+h.c.),
\end{equation}
where
\begin{equation}
  \tilde{E}_{f} = z_{\veck} \bracket{E_{f}+\Sigma_{f}(\veck,0)},
\end{equation}
\begin{equation}
  \tilde{V}_{\veck M\sigma} = \sqrt{z_{\veck}} V_{\veck M\sigma},
\end{equation}
\end{subequations}
where the renormalization amplitude $z_{\veck}$ is defined as
\begin{equation}
  z_{\veck} = \bracket{1-\diff{\Sigma_{f}(\veck,\omega)}{\omega}}^{-1}
                _{\omega=0} \ll 1
\end{equation}
Here, the renormalized $f$-level $\tilde{E}_{f}$ has a {\veck}-dependence
through that of the self-energy $\Sigma_{f}(\veck,0)$ in general. However, we
first investigate the case where the {\veck}-dependence can be neglected,
because the heavy quasiparticles themselves would not be formed if
$\Sigma_{f}(\veck,0)$ had appreciable dispersion. The effect of its
{\veck}-dependence will be discussed later in relation to the resistivity
($\S${\ref{sec:Resistivity}}) and the effect of pressure
($\S${\ref{sec:Pressure}}).

Then we can rewrite the Green functions, eqs.(\ref{eqn:Green}), as
\begin{subequations}
\label{eqn:Green_q}
\begin{equation}
G^{c}_{\sigma}(\veck,\omega)
=\dfrac{1}{\omega-\xi_{k}
-\tilde{V}^{2}_{f}(\veck) / (\omega-\tilde{E}_{f}) }
=\dfrac{A^{c}_{+}(\veck)}{\omega-E^{+}_{\veck}}
+\dfrac{A^{c}_{-}(\veck)}{\omega-E^{-}_{\veck}},
\label{eqn:Gc_q}
\end{equation}
\begin{equation}
G^{f}_{M}(\veck,\omega)
=\dfrac{z_{\veck}}{\omega-\tilde{E}_{f}
-\tilde{V}^{2}_{f}(\veck) / (\omega-\xi_{\veck}) }
=z_{\veck}\parenthesis{\dfrac{A^{f}_{+}(\veck)}{\omega-E^{+}_{\veck}}
+\dfrac{A^{f}_{-}(\veck)}{\omega-E^{-}_{\veck}}},
\label{eqn:Gf_q}
\end{equation}
\end{subequations}
where $E^{\pm}_{\veck}$ are two hybridized-quasiparticle bands given by
\begin{equation}
  E^{\pm}_{\veck}
   = \dfrac{1}{2}\bracket{\xi_{\veck}+\tilde{E}_{f} \pm
       \sqrt{(\xi_{\veck}-\tilde{E}_{f})^{2}+4\tilde{V}^{2}_{f}(\veck)}},
\label{eqn:Ek}
\end{equation}
where
\begin{equation}
  \tilde{V}^{2}_{f}(\veck) = z_{\veck} V^{2}_{f}(\veck).
\label{eqn:V2_q}
\end{equation}
The residues $A^{c}_{\pm}(\veck)$ and $A^{f}_{\pm}(\veck)$ in
eqs.(\ref{eqn:Green_q}) are
\begin{subequations}
\begin{equation}
  A^{c}_{\pm}(\veck) = \bracket{1+\dfrac{\tilde{V}^{2}_{f}(\veck)}
                               {(E^{\pm}_{\veck}-\tilde{E}_{f})^{2}}}^{-1},
\end{equation}
\begin{equation}
  A^{f}_{\pm}(\veck) = \bracket{1+\dfrac{\tilde{V}^{2}_{f}(\veck)}
                               {(E^{\pm}_{\veck}-\xi_{\veck})^{2}}}^{-1},
\end{equation}
\end{subequations}
which give the spectral weight of conduction- and $f$-electrons, respectively,
in the upper/lower bands.

\subsection{Band Insulator of Quasiparticles}

In the case of the electron filling corresponding to the band insulator,
the renormalized Fermi level is located in between $E^{+}_{\veck}$ and
$E^{-}_{\veck}$ forming the hybridization gap. As we have mentioned above,
almost all the compounds called  ``Kondo insulator'' contain even number of
electrons in the unit cell and have a right to be a band insulator.
A difference from the conventional semiconductor is that the hybridization gap
is highly renormalized by strong correlation effect between $f$-electrons.
Hereafter, we investigate the case where the ground crystal-field level is
$\ket{5/2,\pm 3/2}$, which turn out to be consistent with anomalous properties
of CeNiSn as discussed below.~\cite{rf:Crystal} Then, due to
eqs.(\ref{eqn:V2}),(\ref{eqn:Ek}) and (\ref{eqn:V2_q}), the hybridization gap
$\Delta(\veck_{\rm B})$ is given by
\begin{equation}
\Delta(\veck_{\rm B})
\simeq 2\dfrac{\tilde{V}^{2}_{f}(\veck_{\rm B})}{\xi_{\veck_{\rm B}}}
\simeq 2 z_{\veck_{\rm B}}\dfrac{V^{2}_{f}(\veck_{\rm B})}{\xi_{\veck_{\rm B}}}
\equiv T_{\rm K} (1-\hat{k}^{2}_{{\rm B} z})(1+15\hat{k}^{2}_{{\rm B} z})
\label{eqn:gap}
\end{equation}
where $\veck_{\rm B}$ denotes the wavevector at the zone boundary and
$T_{\rm K} \equiv 2z_{\veck_{B}}V^{2}/D$. For simplicity we neglect
$|\veck|$-dependence of $V^{2}$ and {\veck}-dependence of $z_{\veck_{\rm B}}$.
In deriving eq.(\ref{eqn:gap}), we have assumed that the renormalized
hybridization $\sqrt{z_{\veck_{B}}}V$ is much smaller than the bare band-width
of conduction electrons $2D$. Thus the hybridization gap vanishes at points on
the zone boundary where $\hat{k}_{{\rm B} z}=\pm 1$ and becomes a pseudogap.
That is, the renormalized DOS have no clear gap threshold. This is to be
compared to the ``axial-like gap'' in anisotropic superconductors, although
the resultant DOS is totally different as discussed below.

\subsection{Mass Enhancement Factor}

The renormalization amplitude $z$ for the particle-hole symmetric case has
been calculated by Rice-Ueda,~\cite{rf:Rice} on the basis of the Gutzwiller
approximation, and by Shiba~\cite{rf:Shiba}, on the basis of variational Monte
Carlo calculations for the Gutzwiller ansatz, with the use of a model
hybridization $V^{2}(\veck)=V^{2}$. A similar but more extended result has
recently been obtained by numerical renormalization group
method~\cite{rf:Sakai}, quantum Monte Carlo calculation~\cite{rf:Mutou} and
the exact diagonalization method~\cite{rf:Saso} in $d=\infty$ system. We have
performed the calculation similar to Rice-Ueda's with anisotropic
hybridization $V^{2}(\veck)=V^{2}(1-\hat{k}_{z}^{2})$, a simplified version of
(\ref{eqn:V2}). The result for the filling corresponding to band insulator is
\begin{subequations}
\label{eqn:Mass}
\begin{equation}
  z=\dfrac{\e^{19/12}D^{2}}{4V^{2}}\exp\parenthesis{-\dfrac{3UD}{32V^{2}}}
\end{equation}
which is compared with that of Rice-Ueda
\begin{equation}
  z = \dfrac{D^{2}}{V^{2}}\exp\parenthesis{-\dfrac{UD}{8V^{2}}}
\end{equation}
\end{subequations}
where the hybridization gap is constant and fully opened.

In those model calculations, the particle-hole symmetry is assumed, so that
the occupation number of $f$-electron $n_{f}$ per site is exactly unity, i.e.
$n_{f}=1$. However, this constraint is easily relaxed by introducing the
asymmetry of conduction band on the position of the $f$-level. Therefore, it
is possible to calculate the mass enhancement factor in the way similar to
above not only in the Kondo regime, where $n_{f} \simeq 1$, but also in the
valence-fluctuation regime.

\subsection{Effect of Impurity Scattering}
\label{sec:Impurity}

It can be shown, on the basis of the Ward identity arguments, that the
$s$-wave impurity potential $u$ is renormalized by many-body vertex correction
as~\cite{rf:Kotliar}
\begin{equation}
\label{eqn:u}
  u \to \tilde{u}
    =u \cdot \bracket{1-\diff{\Sigma_{f}(\veck,\omega)}{\omega}}_{\omega=0}
    =\dfrac{1}{z}u
\end{equation}
This renormalization is shown in Fig.~1 in terms of the Feynman
diagram. Then, for strongly correlated systems where $z^{-1} = m^{*}/m \gg 1$,
the impurity scattering always becomes that of unitarity limit, i.e.,
$\tilde{u}N_{\rm F} \gg 1$, even if the bare potential $u$ is moderate one,
i.e., $u N_{\rm F} \lsim 1$. Then the Green functions of the conduction- and
$f$-electrons are given by
\begin{subequations}
\label{eqn:Green_imp}
\begin{equation}
  G^{c}_{\sigma}(\veck,\omega)
   = \dfrac{1}{\omega-\xi_{\veck}-\dfrac{\tilde{V}^{2}_{f}(\veck)}
          {\omega-\tilde{E}_{f}-\i z_{\veck}\Im\Sigma_{\rm imp}(\omega)}},
\label{eqn:Gc_imp}
\end{equation}
\begin{equation}
  G^{f}_{M}(\veck,\omega)
   = \dfrac{z_{\veck}}
           {\omega-\tilde{E}_{f}-\i z_{\veck}\Im\Sigma_{\rm imp}(\omega)
    -\dfrac{\tilde{V}^{2}_{f}(\veck)}{\omega-\xi_{\veck}}},
\label{eqn:Gf_imp}
\end{equation}
\end{subequations}
where the self-energy $\Sigma_{\rm imp}(\omega)$ due to impurity scattering is
given in the t-matrix approximation by
\begin{equation}
  \Sigma_{\rm imp}(\omega)
   =n_{\rm imp}\dfrac{\tilde{u}}
                 {1-\tilde{u}{\sum_{\veck}}G^{f}_{M}(\veck,\omega)},
\label{eqn:Sigma}
\end{equation}
where $n_{\rm imp}$ denotes the impurity concentration. In deriving
eqs.(\ref{eqn:Green_imp}), the self-energy of conduction electrons due to
impurity scattering has been neglected because the renormalization, such as
eq.(\ref{eqn:u}), does not occur. Equations (\ref{eqn:Green_imp}) and
(\ref{eqn:Sigma}) need to be solved self-consistently as in the case of heavy
fermion superconductors, where the impurity scattering in the unitarity limit
is known to give rise to appreciable residual DOS in the V-shaped gap even for
a very small impurity concentration.~\cite{rf:Schmitt-Rink,rf:Hirschfeld}

\subsection{Density of States of Quasiparticles}
\label{sec:DOS}

The quasiparticle DOS are calculated as follows:
\begin{subequations}
\label{eqn:DOS}
\begin{eqnarray}
\label{eqn:DOS1}
\tilde{N}(\omega)
&=&\sum_{\veck\sigma}\bracket{\delta(\omega-E^{+}_{\veck})
				     +\delta(\omega-E^{-}_{\veck})}, \\
\label{eqn:DOS2}
&=& N_{\rm F}\int^{1}_{0}d\hat{k}_{z}\int^{D}_{-D}dE
          \parenthesis{1+\dfrac{\tilde{V}^{2}_{f}(\hat{k}_{z})}{E^{2}}}
	  \delta(\omega-E) \theta(|E|-\Delta(\hat{k}_{z})),
\end{eqnarray}
\end{subequations}
where
\begin{equation}
  \Delta(\hat{k}_{z})\equiv T_{\rm K}(1-\hat{k}_{z}^{2})(1+15\hat{k}_{z}^{2}).
\label{eqn:Delta}
\end{equation}

In deriving eq.(\ref{eqn:DOS2}) from eq.(\ref{eqn:DOS1}), we have assumed for
simplicity that the conduction band has a linear dispersion with constant DOS,
$N_{\rm F}$, and extending from $-D$ to $D$, and $\tilde{E}_{f} = 0$. A result
of numerical calculation of ${\tilde N}(\omega)$, eq.(\ref{eqn:DOS2}), is
shown in Fig.~2. Here, the relation between the hybridization $V$ and
renormalization factor $z$ is determined as $z V^{2} / D = 0.01 D$.
The shape of DOS shown in Fig.~2 has two characteristics:
(1) $\tilde{N}(\omega=0)$ is finite, and
(2) it exhibits four-peak structure, i.e., there exist two energy scales
($\Delta_{1}=T_{\rm K}=0.02D,\Delta_{2} \simeq 0.08D$). The former (1)
results from the fact that the hybridization gap vanishes at points
($\hat{k}_{z}= \pm 1$). The shape of DOS around the Fermi level $\omega=0$
can be calculated analytically as
\begin{equation}
\tilde{N}(\omega) \simeq N_{F} \dfrac{D^{2}}{64zV^{2}} \bracket{1+2
  \left(1+\dfrac{19D^{2}}{12\cdot32zV^{2}}\right)\dfrac{\omega}{D}+\cdots}.
\end{equation}
This is in marked contrast with the case of heavy-fermion superconductors,
where the point node leads to DOS proportional to $\omega^{2}$. The reason for
the characteristic (1) to hold is that there exists a singularity
$\propto E^{-2}$, in the first factor of the integrand of (\ref{eqn:DOS2}),
which arises from the Jacobian
$|{\rm d}\xi/{\rm d}E| = 1 + \tilde{V}^{2}_{f}(\hat{k}_{z})/E^{2}$.

The latter characteristic (2) is related to the existence of two extremum
values of $\Delta(\hat{k}_{z})$, eq.(\ref{eqn:Delta}): $\Delta_{1}$ corresponds
to the minimum of eq.(\ref{eqn:Delta}) at $\hat{k}_{z}=0$, at which the
hybridization gap is given by $T_{K}$, and $\Delta_{2}$ corresponds to the
maximum of eq.(\ref{eqn:Delta}) at $\hat{k}_{z}=\sqrt{7/15}$, in which the
hybridization gap becomes maximum. The ratio of $\Delta_{2}$ and $\Delta_{1}$
is given as $\Delta_{2}/\Delta_{1} = 64/15$.

\subsection{Effect of Impurity Scattering on Density of States}
\label{sec:ImpurityDOS}

Next let us consider the effects of impurity scattering on quasiparticle DOS.
From eqs.(\ref{eqn:Gf_imp}) and (\ref{eqn:Sigma}), DOS is calculated
self-consistently as follows:
\begin{subequations}
\label{eqn:DOS_self}
\begin{equation}
  \tilde{N}(\omega) \approx
   \dfrac{1}{\pi}\sum_{\veck}\dfrac{-z_{\veck}\Im\Sigma_{\rm imp}(\omega)}
    {\parenthesis{\omega-\tilde{E}_{f}-\dfrac{z_{\veck}V^{2}_{f}(\veck)}
    {\omega-\xi_{\veck}}}^{2}
    +\bracket{z_{\veck}\Im\Sigma_{\rm imp}(\omega)}^{2}},
\label{eqn:DOS_imp}
\end{equation}
\begin{equation}
  \Im\Sigma_{\rm imp}(\omega)=
   -n_{\rm imp}\dfrac{\tilde{u}^{2}\pi z_{\veck}\tilde{N}(\omega)}
     {1+\tilde{u}^{2}\parenthesis{\pi z_{\veck}\tilde{N}(\omega)}^{2}}.
\label{eqn:Sigma_imp}
\end{equation}
\end{subequations}
Results of numerical solution of eqs.(\ref{eqn:DOS_self}) are shown in
Fig.~3. One can see that the residual DOS at the Fermi level,
$\tilde{N}(\omega=0)$, is very sensitive to the impurity concentration and
drastically increases with the impurity concentration $n_{\rm imp}$.
In Fig.~4 one finds that $\tilde{N}(\omega=0)$ is roughly
proportional to $\sqrt{n_{\rm imp}}$. Precisely speaking, the residual DOS
exists, even if $n_{\rm imp}=0$. So, in the limit, $n_{\rm imp} \to 0$, its
$n_{\rm imp}$-dependence is given by $N_{0}\sqrt{1+c_{0}n_{\rm imp}}$, where
$N_{0}$ is the residual DOS without impurities, and $c_{0}$ is a proper
constant of order unity. This is to be compared with the impurity effects in
heavy fermion superconductors.~\cite{rf:Schmitt-Rink,rf:Hirschfeld}
The square-root dependence of the residual DOS has also been derived on the
basis of a different picture, where it is understood as an impurity band
similar to the doped semiconductor; namely the doping accompanied by variation
of carrier number is necessary to obtain finite DOS in the true
gap.~\cite{rf:Schlottmann,rf:Shiina} Our theory has been developed to discuss
the case where carrier number does not change, while it is easily extended to
the case where carriers are doped.

\section{Physical Properties}
\label{sec:Physical}

In this section, we study the qualitative aspects of several physical
quantities and compare them with experiments.

\subsection{Specific Heat}
\label{sec:SpecificHeat}

The specific heat is calculated on the basis of the quasiparticle picture,
and the electronic specific heat coefficient, $\gamma \equiv C(T)/T$, is
given as follows:
\begin{equation}
  \gamma = 2 \int_{0}^{\infty} dx \tilde{N}(E) x^{2} \sech^{2}(x),
\end{equation}
where $x=E/2T$. The temperature dependence of $\gamma$ is calculated with
the use of eqs.(\ref{eqn:DOS}) and (\ref{eqn:DOS_self}) and is shown in
Fig.~5. A peak structure is found at $T \sim \Delta_{1}/2$. It is
noted that $\gamma$ is finite at $T=0$, which results from the existence of
the residual DOS at $n_{\rm imp}=0$. If $\Delta_{1}$ is fixed as
$\Delta_{1}/2=7$K, these results are in good agreement with the experimental
data~\cite{rf:Nakamoto,rf:Nishigori,rf:Suderow} at $T < \Delta_{1}$, where the
maximum of theoretical curve for $\gamma$ is adjusted so as to agree with the
experiment of ref.~\ref{rf:4}. We have also verified that the same quality of
agreement with the data of ref.~\ref{rf:5} is obtained while its absolute value
of $C_{m}/T$ is about $10\%$ larger than that of ref. \ref{rf:4}. However at
$T > \Delta_{1}$ the agreement become poor. It is partly improved by
considering temperature dependence of the quasiparticle
DOS.~\cite{rf:Nishigori}

Futhermore, assuming that the effect of the magnetic field is only inducing
the Zeeman splitting, the magnetic-field dependence of $\gamma$ is obtained as
\begin{equation}
  \gamma = \int_{0}^{\infty} dx \tilde{N}(E)
      \bracket{x^{2}_{+}\sech^{2}(x^{2}_{+})+x^{2}_{-}\sech^{2}(x^{2}_{-})},
\end{equation}
where $x_{\pm} = (E \pm h)/2T$, $h = g_{J}\mu_{\rm B}|J_{z}|H$, $g_{J}$ being
a g-factor. The results for various temperatures are shown in Fig.~6.
The coefficient $\gamma$ at low temperature exhibits two-peak structure, which
reflects the peak structure of DOS. This prediction has not yet been observed,
partly because the strength of the magnetic field is not enough.

\subsection{Longitudinal Relaxation Rate of NMR}
\label{sec:NMR}

The longitudinal NMR relaxation rate, $1/T_{1}$, is obtained as follows
\begin{subequations}
\begin{eqnarray}
\dfrac{1}{T_{1}T}
&\propto& \lim_{\omega \to 0}
            \sum_{\bf q}\dfrac{\Im\chi^{-+}({\bf q},\omega)}{\omega}, \\
\label{eqn:1/T1T}
&\propto& \int_{0}^{\infty} dE
            \tilde{N}(E)^{2}\sech^{2}(\dfrac{E}{2T})\dfrac{1}{T}.
\end{eqnarray}
\end{subequations}
Here we have assumed that the quasiparticle DOS directly affects $1/T_{1}$ at
Sn site via the $c$-$f$ exchange as done in the analysis of experimental data.
The results of numerical calculations of (\ref{eqn:1/T1T}), together with
experimental data,~\cite{rf:Nakamura} are shown in Fig.~7 for the
same parameters as in Fig.~5.
For the temperature region,
$0.1 \Delta_{1} \lsim T \lsim \Delta_{1} \simeq 0.02 D$, $1/T_{1}$ shows
the $T^{3}$-like behavior reflecting the formation of the pseudogap at
$\omega \simeq \Delta_{1}/2$ of DOS as shown in Fig.~2, and for
$T \lsim 0.1\Delta_{1}$, it shows $T$-linear behavior reflecting the residual
DOS at the Fermi level. These behaviors well reproduce the $T$-dependence of
$1/T_{1}$ observed in the experiments,~\cite{rf:Nakamura} if $\Delta_{1}$ is
fixed as $\Delta_{1}=14$K. In addition, as increasing impurity concentration
$n_{\rm imp}$, the residual DOS rises up drastically and the $T$-linear
behavior masks the $T^{3}$-like behavior. These are also in agreement with
the experiments,~\cite{rf:Nakamura} where the residual DOS shows the
$n_{\rm imp}$-dependence quite similar to the theoretical curve shown in
Fig.~4.

The above results have been derived on the basis of the quasiparticle picture,
so that, strictly speaking, its validity is assured only in the low temperature
region $T<T_{\rm K}$. However, it may be extended to much higher temperature
region as far as the qualitative aspects are concerned. Indeed, $1/T_{1}$ in
Fig.~7 exhibits the localized character of $f$-electrons for
$T>T_{\rm K}=0.02 D$, $1/T_{1}\propto T^{0}$, and approaches asymptotically to
the Korringa-like behavior at much higher temperatures, $1/T_{1}\propto T$, as
can be inferred from the DOS of quasiparticles shown in Fig.~2.
In the latter region, the NMR relaxation is expected to occur mainly through
the coupling with the conduction electrons as in LaNiSn.~\cite{rf:Kyogaku}
Recently such a behavior has been recognized by an analysis of the data of
1/$T_{1}$ up to the room temperature.~\cite{rf:Ohama}

\subsection{Magnetization}
\label{sec:Magnetization}

The quasiparticle contribution to the magnetization is calculated as
\begin{equation}
  M=\int_{-\infty}^{\infty} dE \bracket{f(E-h)-f(E+h)}\dfrac{\tilde{N}(E)}{2}.
\end{equation}
The results are shown in Fig.~8 where one can see that the
magnetization, in the unit $g_{J}\mu_{\rm B}|J_{z}|$, is proportional to $h$,
in the unit $D$, at low magnetic field and the slope is given by the residual
DOS.
However, as increasing $h$ the magnetization drastically increases at $h>0.01D$
owing to the two large humps of DOS at $\omega=\pm\Delta_{1}/2$.
These tendencies are found in the experimental data~\cite{rf:Sugiyama}.
The slope of the magnetization at low field is enhanced by small amount of
impurity and the whole structures of $M$-$h$ curve shade off.

\subsection{Uniform Spin Susceptibility}
\label{sec:Susceptibility}

The uniform spin susceptibility along the easily axis ($a$-axis), in the unit
$(g_{J}\mu_{\rm B}|J_{z}|)^{2}/D$, is given by the derivative of
the magnetization as
\begin{equation}
\label{eqn:chi00}
\Re\chi(0,0) = \left.\diff{M}{h}\right|_{h \to 0}
  = \int_{-\infty}^{\infty} dE \bracket{-\diff{f(E)}{E}}\tilde{N}(E).
\end{equation}
Its temperature dependence shown in Fig.~9 exhibits the peak
structure like $\gamma$ as discussed in $\S${\ref{sec:SpecificHeat}}.
However, the temperature at which $\Re\chi(0,0)$ has the maximum
($T \simeq \Delta_{1}$) is different from that for $\gamma$. Increasing
impurity concentration, the sharp dip at low temperature is filled up rapidly.
Since the susceptibility (\ref{eqn:chi00}) is given only by the contribution of
the quasiparticles, the Van Vleck term is not included. If the latter is simply
a constant, the observed Knight shift~\cite{rf:Nakamura} for $H \| a$
represents the behavior of the uniform spin susceptibility $\chi^{a}$, which
is in agreement with the curves of Fig.~9.

\subsection{Neutron Scattering}
\label{sec:Neutron}

The imaginary part of the dynamical susceptibility, the spectral weight of spin
fluctuations, is calculated at $T = 0$ without a vertex correction as
\begin{equation}
\label{eqn:chi}
  \Im\chi({\bf Q},\omega)\simeq
    \pi\sum_{\veck}\bracket{f(E_{\veck}^{-})-f(E_{\bf k+Q}^{+})}
			 \delta(\omega-E_{\bf k+Q}^{+}+E_{\veck}^{-})
\end{equation}
where $f(E)$ is the Fermi distribution function and $f(E_{\veck}^{-})=1$ and
$f(E_{\veck}^{+})=0$ at $T \to 0$. The spectral weight (\ref{eqn:chi}) has been
calculated numerically with the use of the quasiparticle dispersion,
eq.(\ref{eqn:Ek}), for specified ${\bf Q}$'s.
The spectral weight at ${\bf Q}=(1/2,0,0)$ is shown in Fig.~10 (a),
which shows that there exists a broad hump at around $\omega=0.08D$. (It is
noted that $x$-, $y$-, and $z$-axis here corresponds to the $b$-, $c$-, and
$a$-axis, respectively of CeNiSn; so that ${\bf Q}=(1/2,0,0)$ implies
${\bf Q}=[0,1/2,0]$ in the notation of experiments of CeNiSn for instance.)
This structure corresponds to the transition from one peak of DOS at
$\omega=-0.04D$ to another at $\omega=0.04D$ in DOS. For example, the former
peak corresponds to the quasiparticle at $\veck_{1}=(-1/4,1/2,1/2)$, while the
latter at $\veck_{2}=(1/4,1/2,1/2)$, because
$\hat{k}_{1z} = \hat{k}_{2z} \sim \sqrt{7/15}$.
Thus, ${\bf Q}=\veck_{2}-\veck_{1}=(1/2,0,0)$. The spectrum at
${\bf Q}=(0,0,1/2)$ shown in Fig.~10 (b) has a peak at around
$\omega=0.06D$, which corresponds to the energy from an edge of the gap at
$\omega=-0.02D$ to one peak at $\omega=0.04D$ in DOS. For example, the former
peak corresponds to the quasiparticle at $\veck_{3}=(1/2,0,0)$, while the
latter at $\veck_{4}=(1/2,0,1/2)$, because $\hat{k}_{3z}=0$ and
$\hat{k}_{4z} \sim \sqrt{7/15}$. Thus, ${\bf Q}=\veck_{4}-\veck_{3}=(0,0,1/2)$.
It is seen that a difference between the two spectra arises from that of the
possibility of the zero energy transition. Since the hybridization gap vanishes
along the $z (a)$-axis, there exist zero energy excitations for the transition
between the points on the $z (a)$-axis. This is because we make choice of
non-dispersive $f$-level. However, generally speaking, $\Im\chi({\bf Q},0)$
have to vanish in the normal Fermi liquid theory by the symmetry reason.
This anomaly is removed by taking into account the small dispersion of
$f$-electrons due to a possible weak $\veck$-dependence of the $f$-electron
self energy, $\Sigma_{f}(\veck,0)$.

The spectral intensity at $\omega=0.08D \simeq \Delta_{2}$ and
${\bf Q}=(Q_{x},0,0)=[0,Q_{b},0]$ is computed as a function of
$Q_{x}$($Q_{b}$) and shown in Fig.~11. One can see the peak at
$Q_{x}(Q_{b})=1/2$. The reason is that the peak shifts to higher energy as
deviating from $Q_{x}(Q_{b})=1/2$. These features are consistent with
$Q_{b}$-dependence of the intensity at 4.25meV in the inelastic neutron
scattering.~\cite{rf:Kadowaki}

The details of these spectra are modified according to choices of the
dispersions of conduction electrons, though characteristic structures do not
change. However, when the conduction band do not cross the $f$-level in some
direction in the $\veck$-space, the excitation energy is rather higher and the
spectrum can not be observed at low energy region in general. This case may be
realized in the low-energy spectrum at ${\bf Q} \parallel c$-axis in the
experiment.~\cite{rf:Kadowaki}

\subsection{Anisotropy of Resistivity}
\label{sec:Resistivity}

It is the resistivity that is one of the measures to classify the heavy-fermion
materials into  ``Kondo insulator'' or not. The resistivity in heavy fermions
exhibits the Kondo effect at high temperature region and metallic or
activation-type behavior at low temperature region. We are interested in the
behavior at temperatures lower than the coherent temperature $T_{\rm coh}$,
in which the current is carried by the quasiparticles. In this case the
conductivity can be evaluated by
\begin{equation}
  \sigma_{\mu\nu}(T) \propto
    \sum_{\veck}J_{{\veck}\mu}J_{\veck\nu}\tau_{\veck}
      \parenthesis{-\diff{f(E_{\veck})}{E_{\veck}}},
\end{equation}
where $J_{\veck\mu}$ is the velocity of the quasiparticle and $\tau_{\veck}$ is
its lifetime.~\cite{rf:Yamada2} Assuming that the microscopic expression of the
current is given only by the conduction electrons (neglecting the dispersion of
$f$-electrons), the anisotropy of the conductivity is given as follows:
\begin{subequations}
\label{eqn:sigma}
\begin{eqnarray}
\sigma_{\|}
& \propto & \sum_{\veck}A_{\pm}^{c}(\veck)^{2}v_{z}(\veck)^{2}
                \tau_{\veck}\parenthesis{-\diff{f(E^{\pm}_{\veck})}
		                        {E^{\pm}_{\veck}}}, \\
\label{eqn:sigma_p}
& \propto &
     \int dE_{\veck} \int d\Omega_{\veck} \dfrac{E_{\veck}^{2}}
                 {\parenthesis{-z_{\veck}\Im\Sigma_{f}(\veck,E_{\veck})}
		  \tilde{V}_{f}^{2}(\veck)
		  +\parenthesis{-\Im\Sigma_{c}(\veck,E_{\veck})E_{\veck}^{2}}}
		\parenthesis{-\diff{f(E_{\veck})}{E_{\veck}}}, \\
\label{eqn:sigma2}
\sigma_{\bot}
& \propto & \sum_{\veck}A_{\pm}^{c}(\veck)^{2}v_{x}(\veck)^{2}
                \tau_{\veck}\parenthesis{-\diff{f(E^{\pm}_{\veck})}
		                        {E^{\pm}_{\veck}}}, \\
\label{eqn:sigma_b}
& \propto & \int dE_{\veck} \int d\Omega_{\veck}
                 \dfrac{\sin^{2}\theta_{\veck}E_{\veck}^{2}}
		       {\parenthesis{-z_{\veck}\Im\Sigma_{f}(\veck,E_{\veck})}
			\tilde{V}_{f}^{2}(\veck)
		 +\parenthesis{-\Im\Sigma_{c}(\veck,E_{\veck})E_{\veck}^{2}}}
		  \parenthesis{-\diff{f(E_{\veck})}{E_{\veck}}},
\end{eqnarray}
\end{subequations}
where $\sigma_{\|}$ and $\sigma_{\bot}$ are the conductivity along the
$a$-axis and in the $bc$-plane, respectively, and
\begin{subequations}
\begin{equation}
\label{eqn:lifetime}
  \dfrac{1}{\tau_{\veck}}=\dfrac{1}{E_{\veck}^{2}+\tilde{V}_{f}^{2}(\veck)}
    \bracket{\parenthesis{-z_{\veck}\Im\Sigma_{f}(\veck,E_{\veck})}
    \tilde{V}_{f}^{2}(\veck)+
    \parenthesis{-\Im\Sigma_{c}(\veck,E_{\veck})E_{\veck}^{2}}},
\end{equation}
\begin{equation}
  v_{z}(\veck)=\diff{\xi_{\veck}}{k_{z}} \ \ {\rm and} \ \
  v_{x}(\veck)=\diff{\xi_{\veck}}{k_{x}}.
\end{equation}
\end{subequations}
Here we have introduced $\Sigma_{c}(\veck,E_{\veck})$, the self-energy of the
conduction electrons, because $\sigma_{\|}$ in pure system diverges
logarithmically otherwise, reflecting the fact that the conduction electrons
are decoupled from $f$-electrons in the $z$-direction ($a$-axis) where the
hybridization vanishes. We regard the renormalization factor of conduction
electrons from $\Sigma_{c}(\veck,E_{\veck})$ as 1. In this model the
conductivity in the $bc$-plane is isotropic unless the anisotropy of the
conduction band is taken into account. In deriving (\ref{eqn:sigma_b}) from
(\ref{eqn:sigma2}), we have taken into account only the low energy excitations
around $\hat{k}_{z}=\pm 1$ so that the obtained result should be regarded as
that for asymptotic behavior in the limit $T \to 0$.

In order to discuss the temperature dependence of the conductivity at low
temperature region, we must calculate the energy dependence of the imaginary
part of the self-energies. For simpicity we calculate these along a standard
treatment of the Fermi liquid theory neglecting the momentum dependence of the
full vertex:
\begin{equation}
  \Im\Sigma_{\mu}(\veck, E_{\veck}^{+})  \propto
     \int d{\bf p} d{\bf q}
        A_{+}^{\mu}({\bf k-q})A_{-}^{\mu}({\bf p})A_{+}^{\mu}({\bf p+q})
        \delta(E_{\veck}^{+}+E_{\bf p}^{-}-E_{\bf k-q}^{+}-E_{\bf p+q}^{+}),
\end{equation}
where $\mu = c$ or $f$. These integrations are computed by the Monte Carlo
calculation. The numerical results are shown in Fig.~12.
From this one can see that the energy dependence of
$\Im\Sigma_{c}(\veck,E_{\veck})$ and $\Im\Sigma_{f}(\veck,E_{\veck})$ near
the Fermi level at zero temperature can be approximated by $E_{\veck}^{5}$ and
$E_{\veck}^{3}$, respectively. Futhermore, we assume that the temperature
dependence of $\Im\Sigma_{\mu}(\veck, E_{\veck})$ is given with replacing
$E_{\veck}^{2}$ by max ($E_{\veck}^{2}$ and $(\pi T)^{2}$), as can be seen from
the structure of the Green functions. By using these results we can estimate
the temperature dependence of the resistivity. Substituting these energy
dependence into eqs.(\ref{eqn:sigma}), we obtain up to the logarithmic accuracy
\begin{subequations}
\begin{equation}
  \sigma_{\|} \propto T^{-1} \ \ ,\ {\rm and} \ \ \sigma_{\bot} \propto T^{0},
\end{equation}
i.e,
\begin{equation}
  \rho_{\|} \propto T \ \ ,\ {\rm and} \ \ \rho_{\bot} \propto T^{0}.
\end{equation}
\end{subequations}

Next we discuss the effect of impurity scattering on $\sigma_{\|}$.
For simplicity we first assume that the self-energy
$\Im\Sigma_{c}(\veck,E_{\veck})$ and $\Im\Sigma_{f}(\veck,E_{\veck})$ are
independent of $E_{\veck}$ and proportional to the impurity concentration
$n_{\rm imp}$. In this case, by using eq.(\ref{eqn:sigma_p}), we obtain
\begin{equation}
\label{eqn:sigma_imp}
  \sigma_{\|}^{\rm imp} \propto T^{2}/n_{\rm imp} \ \ ,\ {\rm i.e.} \ \
  \rho_{\|}^{\rm imp}   \propto n_{\rm imp}T^{-2}.
\end{equation}
 However, if the small dispersion of $f$-electrons due to a possible weak
$\veck$-dependence of the $f$-electron self energy, $\Sigma_{f}(\veck,0)$,
the current can be carried also by $f$-electrons, so that
\begin{subequations}
\begin{eqnarray}
\sigma_{\|}
& \propto & \int dE_{\veck} \int d\Omega_{\veck}
    \dfrac{1}{E_{\veck}^{2}} \dfrac{\tilde{V}_{f}^{4}(\veck)}
     {\parenthesis{-z_{\veck}\Im\Sigma_{f}(\veck,E_{\veck})}
                   \tilde{V}_{f}^{2}(\veck)
     +\parenthesis{-\Im\Sigma_{c}(\veck,E_{\veck})E_{\veck}^{2}}}
		\parenthesis{-\diff{f(E_{\veck})}{E_{\veck}}}, \\
& \propto & \dfrac{1}{n_{\rm imp}}.
\end{eqnarray}
\end{subequations}
Thus the singularity of the resistivity (\ref{eqn:sigma_imp}) at $T = 0$ is
suppressed. Nevertheless, the residual resistivity in the limit $T \to 0$
increases drastically as increasing the impurity concentration $n_{\rm imp}$.
If we take all the contributions of the quasiparticles into consideration,
including the logarithmic corrections, the resistivity is given by
\begin{equation}
\label{eqn:rho}
  \rho_{\|}
   = \dfrac{\rho_{0}}{\parenthesis{n_{\rm imp}+(T/T_{\rm K})^{3}}^{-1}c_{1}+
       \parenthesis{\dfrac{T/T_{\rm K}}{\log(c_{2}T_{\rm K}/T)}+
       \dfrac{n_{\rm imp}}{(T/T_{\rm K})^{2}\log(c_{3}T_{\rm K}/T)} }^{-1}},
\end{equation}
where $c_{1} (\sim 0.1)$, $c_{2} (\sim 10)$ and $c_{3} (\sim 5)$ are fitting
parameters, which are connected with the small dispersion of $f$-electrons,
the interaction between conduction electrons and impurity scattering of
conduction electrons with the Born approximation, respectively.
$n_{\rm imp}\rho_{0}/c_{1}$ is the resistivity at $T \to 0$. The resistivity
for proper parameters are shown in Fig.~13. These results are in
qualitative agreement with the experimental
data.~\cite{rf:Takabatake2,rf:Nakamoto}
In particular, the temperature dependence of $\rho_{\|} (\rho_{a})$ observed
in the best sample to date is well reproduced as seen in Fig.~13.
It is also found that the resistivity is sensitive to the concentration of
impurities at low temperature in consistent with the experiments.

It should be remarked here that the weak $\veck$-dependence of
$\Sigma_{f}(\veck,0)$ inevitably gives rise to a small semimetallic Fermi
surface around $X$-point (intersection of $a$-axis and the zone boundary) in
general, so long as the hybridization vanishes along the $a$-axis as
eq.(\ref{eqn:V2}). However, such a small Fermi surface is expected to give
only little effect on the qualitative behavior of DOS of quasiparticles
discussed in previous sections, while it sensitively affects the low
temperature behavior of the resistivity especially in the case where the
impurity scattering greatly enhances the resistivity as in eq.
(\ref{eqn:sigma_imp}) when there exists no $\veck$-dependence of
$\Sigma_{f}(\veck,0)$. In deriving (\ref{eqn:rho}), we have taken into account
the dispersion of $f$-electron through the $\veck$-dependence of
$\Sigma_{f}(\veck,0)$, nevertheless we have used the same DOS as
eq.(\ref{eqn:DOS}). In this sense, the calculation is not self-consistent and
the resistivity (\ref{eqn:rho}) should be regarded as a provisional one.
However, the expression (\ref{eqn:rho}) gives a good description for
$T \sim \Delta_{1}/2$, or a good starting point at least.

We should have a few words about $\rho_{\bot}$. The resistivity $\rho_{b}$ and
$\rho_{c}$ of the same sample as shown in Fig.~13 exhibits a dip at
$T \sim 3$K and saturate at $T \to 0$.
This may be understood as follows. Since quasiparticles around $\hat{k}_{z}=0$
with excitation energy ($\sim \Delta_{1}/2$) have large dispersion along $x$-
or $y$-direstion, those are expected to contribute considerably to the
conduction perpendicular to $z(a)$-axis at $T \sim \Delta_{1}/2$, leading to
the suppression of $\rho_{\bot}$ ($\rho_{b}$ and $\rho_{c}$). If we use
$\Delta_{1}/2 \simeq 7$K estimated above by means of $1/T_{1}$'s result,
such suppression or dip is expected to occur at $T \lsim 7$K in consistent
with the above observation.

Now we briefly discuss about the longitudinal magnetoresistance (MR).
The change from the positive MR to the negative MR is observed with increasing
magnetic field.~\cite{rf:Takabatake2,rf:Nakamoto} The negative MR at higher
field region may be caused by the suppression of spin fluctuations by magnetic
field such as in the impurity Kondo effect. However, the case of the positive
MR at low field is more complicated. We have calculated magnetic-field
dependence of the lifetime $\tau_{\veck}$ of quasiparticles and verified that
$\tau_{\veck}$ is a decreasing function of the magnetic field in the low-field
region. Although DOS at low energies increases with the magnetic field due to
the Zeeman splitting of DOS, shown in Fig.~2, the weight of
conduction electrons of those states decreases in general leading to the
enhancement of the resistivity. Therefore, MR is determined on such a delicate
balance between the effects on the lifetime of the quasiparticles and the
details of DOS at the Fermi level. The experimental data can be understood as
follows: MR at lower fields is positive by shortening of the lifetime and at
higher fields becomes negative due to the drastic increment of quasiparticles
which can carry the current.

\subsection{Pressure Dependence}
\label{sec:Pressure}

Next we discuss the pressure dependence on the property of quasiparticles.
As the lattice constant becomes short under the pressure, both the band-width
$D$ of conduction electrons and the hybridization $V$ are enlarged.
However, the fundamental energy scale $V^{2}/D$ is expected to be an increasing
function of the pressure because $V$ is much more sensitive than $D$ for
heavy fermions where $V$ arises through rather small overlap between $f$- and
conduction electrons. Much more pronounced effect of pressure on the
hybridization gap (\ref{eqn:gap}) arises through enlargement of the
renormalization amplitude $z_{\veck}$. This is because $z_{\veck}$ is an
exponentially small qauntity as (\ref{eqn:Mass}) for heavy fermions so that its
relative change under pressure is far larger than that for $V$ and $D$
themselves. Therefore, the energy scale of the gap is expected to increase by
applying the pressure, and so is the resistivity in eq.(\ref{eqn:rho}), which
is scaled by $T_{\rm K}$.

This tendency is consistent with the behavior of CeNiSn and CeRhSb
where the peak of the resistivity shifts to the high temperature by
the pressure.~\cite{rf:Kurisu,rf:Uwatoko} And also the suppression of those
resistivity with the pressure in the limit $T \to 0$ can be understood as
follows. As discussed in $\S${\ref{sec:Resistivity}}, there exists a very tiny
semimetallic Fermi surface at around $\veck=(0,0,\pm 1)$ in general,
so long as $\Sigma_{f}(\veck,0)$ has the dispersion along the $a$-axis
no matter how small it is. After the Fermi surface grows further under
the pressure due to the growth of the dispersion of $\Sigma_{f}(\veck,0)$,
an apparent semimetallic behavior is expected to prevail leading to
the suppression of the resistivity. That is to say, a parameter $c_{1}$,
which corresponds to $\partial \Sigma_{f}(\veck,0) / \partial \veck$,
increases with the pressure in the formula (\ref{eqn:rho}), so that
the resistivity in the limit $T \to 0$, $n_{\rm imp}\rho_{0}/c_{1}$,
is suppressed.

\subsection{Lifetime of Quasiparticles}
\label{sec:Lifetime}

We have neglected an effect of quasiparticle damping due to inelastic
scattering so far.  Here we briefly discuss its effect on DOS of
quasiparticles and temperature dependence of physical quantities.

According to eq. (\ref{eqn:lifetime}), the lifetime of quasiparticles around
$\hat{k}_{z}=\pm 1$ is nearly proportional to
$(E_{\veck}^{2}+\tilde{V}_{f}^{2}(\veck))/
\tilde{V}_{f}^{2}(\veck)E_{\veck}^{3}$ at low energy region
$E_{\veck}<\Delta_{1}$. This lifetime is longer than in the case of the normal
Fermi liquid theory for $E_{\veck} < \Delta_{1}$. This is because the
scattering between quasiparticles is suppressed at low energy, owing to the
restriction of phase space satisfying the energy-momentum conservation law.
Namely, the quasiparticles with low energy are located around (0,0,$\pm 1$),
so that such phase space is restricted within narrow region around
(0,0,$\pm 1$).  These quasiparticles make the flat part near the Fermi level
at DOS of Fig.~2.

The quasiparticle, corresponding to the peak structure at
$\omega = \Delta_{1}/2$ in DOS of Fig.~2, are located along
$\hat{k}_{z}=0$. These quasiparticles also suffers little inelastic
scatterings, again because of the restriction due to the energy-momentum
conservation law.  Thus these lifetime is very long, leading to
$\Im \Sigma_{f}(\hat{k}_{z} \sim 0,E_{\veck}) \sim 0$.  So it is expected that
the peak structure at $\omega = \Delta_{1}/2$ in DOS remains even if the effect
of inelastic scattering is taken into account.

On the contrary, we have no reason to keep the sharp peak structure at
$\omega = \Delta_{2}/2$ in DOS, because the restiction due to the
energy-momentum conservation does not suppress the inelstic scattering of
quasiparticles forming this peak. So the structure at $\omega = \Delta_{2}/2$
probably becomes a broad hump.

However, we believe that the two peak structure in DOS of Fig.~2
remains even though the effect of inelastic scattering is taken into account.
Indeed, the DOS calculated by the second order perturbation theory exhibits two
peak structure similar to those of Fig.~2, although the peak at
$\omega = \Delta_{2}/2$ is somewhat broadened. This result will be discussed
elsewhere.

Futhermore, $T>\Delta_{1}$ is the temperature region where the damping effect
of the quasiparticles affects temperature dependence of physical quantities.
For $T>\Delta_{1}$, these peak structures of DOS fade out,
while the physical quantities are averaged by temperature dependence of
the Fermi distribution. Therefore, the neglect of the damping effect may
give rise to no serious errors as far as the qualitative temperature
dependence is concerned.

\section{Summary and Discussions}

On the basis of the idea of ``adiabatic continuity'', a theory of the
anisotropic semiconductor of heavy fermions has been developed to explain the
anomalous properties of CeNiSn and its isostructural compounds. A difference
from the conventional semiconductors is that the band gap is formed by the
highly renormalized quasiparticles near the Fermi level. So the gap has meaning
only at low temperature region $T<\Delta_{1}$ (corresponding to the
hybridization gap), while the coherent peak of quasiparticles fades out
exhibiting the behaviors of the Kondo lattice metals.~\cite{rf:Ekino}

Wide range of anomalies of CeNiSn can be understood by a model of the
anisotropic hybridization gap which vanishes along the $a$-axis.
The anisotropy of the gap reflects a $\veck$-dependence of hybridization matrix
elements between the conduction electrons and the $f$-electron with particular
symmetry of the crystal field state. The desirable $\veck$-dependence occurs
if the lowest crystal field state consists mainly of $\ket{5/2,\pm3/2}$ due to
its approximately trigonal symmetry and the conduction electrons near the Fermi
level are described by the plane waves, as discussed in
\S {\ref{sec:Hamiltonian}} and \S {\ref{sec:Hybridization}}.

Since there is no band calculation of LaNiSn available to date, it is difficult
to assess whether the latter condition is fulfilled in CeNiSn. However, it may
be not unrealistic to assume that the state of conduction electrons hybridizing
with the $f$-electron localized at Ce site can be approximated by the plane
waves with the wave vector $|$\veck$+{\bf G}| \ <$ several $\times$ (2$\pi/a$),
{\bf G} being some reciprocal lattice vector and $a$ being the lattice
constant. This is because the only way for the $f$-electron to mix with
electrons on different sites is through the mixing with the plane wave states
outside the muffin-tin spheres so long as the conventional LAPW calculation is
performed.

Band calculations of CeNiSn shows that the bands around the Fermi level have
mainly Ce $4f$ character with mixture of Ni 3$d$
component.~\cite{rf:Yanase,rf:Hammond}
So, in the tight-binding picture, the hybridization is expected
to arise through the overlap of Ce $4f$ wavefunction and tails of Ni $3d$
wavefunction.  It is seen by a simple calculation of the tight-binding model
that the hybridization between $f$-electron in the state $\ket{5/2,\pm3/2}$ and
$d$-electrons on the surrounding ions with trigonal symmetry vanishes on the
$k_{z}$ axis, i.e., $V(0,0,k_{z})=0$.

In order to obtain more solid picutre of the $\veck$-dependence of the
hybridization, we need more information of the band structure of LaNiSn.
It is also interesting to discuss a difference between CeNiSn and the so-called
``Kondo insulator", such as Ce$_{3}$Bi$_{4}$Pt$_{3}$~\cite{rf:Thompson} and
YbB$_{12}$~\cite{rf:Kasaya}, with non-vanishing gap in any directions of the
Brillouin zone. From the present point of view, its difference is attributed to
that of the $\veck$-dependence reflecting the symmetry of the lowest
crystal-field level. We leave such discussions for future studies.

\section*{Acknowledgements}
We would like to acknowledge Y. Kitaoka and K. Nakamura for leading to our
attentions to this problem and stimulating discussions. We have much
benefitted from informative conversations and correspondences with
T. Takabatake, S. Nishigori, T. Ekino, Y. Ishikawa, J. Sakurai, K. Sugiyama,
H. Kadowaki, and T. Sato. One of the authors (K. M.) would like to acknowledge
J. Flouquet, S. Kambe, H. Suderow, and S. Raymond for valuable discussions and
hospitality at CENG. K. M. also acknowledges H. Harima and A. Yanase for
providing him with the unpublished figure of the Fermi surface of CeNiSn due to
their band structure calculations. One of the authors (H. I.) is grateful to
Y. Ohashi, T. Mutou and G. Nakamoto for useful discussions.
This work is supported by the Grant-in-Aid for Scientific Research (07640477),
and Monbusho International Scientific Program (06044135), and the Grant-in-Aid
for Scientific Research on Priority Areas ``Physics of Strongly Correlated
Conductors'' (06244104) of Ministry of Education, Science and Culture.
This work is also supported in part by The Japanese-German Cooperative
Science Promotion Program of Japan Society for Promotion of Science.

\newpage

\footnotesize
\begin{description}
\item[Fig.~1 \ \ ]
The diagram for the many-body vertex correction of impurity
scattering. The broken line represents the impurity potential $u$ of $s$-wave,
the external solid line the Green function of the $f$-electrons, and internal
solid line stands both $f$-electrons and conduction electrons. $\Gamma$ is the
full vertex due to the Coulomb repulsion between $f$-electrons. $\tilde{u}$ is
a renormalized potential of impurity scattering.
\item[Fig.~2 \ \ ]
The density of states, $\tilde{N}(\omega)$, of the quasiparticles as
a function of $\omega$ in the unit $D$, half the band-width of conduction band.
The Fermi level is located at $\omega=0$.
\item[Fig.~3 \ \ ]
The effect of the impurity scattering on the density of states.
\item[Fig.~4 \ \ ]
The density of states at the Fermi level as a function of the
impurity concentration.
\item[Fig.~5 \ \ ]
The temprature dependence of the specific heat coefficient $\gamma$.
The unit of the temperature is $D$, half the band-width of conduction
electrons. Circles are experimental data ($C_{\rm m}/T$) of ref.~\ref{rf:4}.
$\Delta_{1}/2=0.01D$ corresponds to 7K.
\item[Fig.~6 \ \ ]
The specific heat coefficient, $\gamma$, as a function of the magnetic
field, $h$. The unit of $h$ is $D$, half the band-width of conduction
electrons.
\item[Fig.~7 \ \ ]
The longitudinal NMR relaxation rate, $1/T_{1}$ (in arbitrary unit),
as a function of the temprerature $T$ in the unit $D$, half the band-width of
conduction electrons. Triangles and crosses are experimental data of ref.
\ref{rf:3}. $\Delta_{1}=0.02D$ corresponds to 14K.
\item[Fig.~8 \ \ ]
The magnetization, $M$ (in the unit $g_{J}\mu_{\rm B}|J_{z}|$), as a
function of the magnetic field $h$ (in the unit $D$) at $T=0$.
\item[Fig.~9 \ \ ]
$T$-dependence of the uniform susceptibility, $\Re \chi(0,0)$ in the
unit $(g_{J}\mu_{\rm B}|J_{z}|)^{2}/D$. The unit of $T$ is $D$, half the
band-width of conduction electrons.
\item[Fig.~10 \ \ ]
The spectral weight of spin fluctuations, $\Im \chi({\bf Q},\omega)$
(in arbitrary unit), as a function of $\omega$ at $T=0$ for
(a) ${\bf Q}=(1/2,0,0)=[0,1/2,0]$ and (b) ${\bf Q}=(0,0,1/2)=[1/2,0,0]$.
$\Im\chi({\bf Q},0)$ is finite in (b) due to a choice of non-dispersive
$f$-level.
\item[Fig.~11 \ \ ]
The intensity of $\Im \chi({\bf Q},\omega=\Delta_{2})$ in arbitrary
unit as a function of $Q_{x}$ ($Q_{b}$) of ${\bf Q}=(Q_{x},0,0)=[0,Q_{b},0]$
at $T=0$.
\item[Fig.~12 \ \ ]
The energy dependence of the self-energies,
(a) $\Im\Sigma_{c}(\veck,E_{\veck})$ and (b) $\Im\Sigma_{f}(\veck,E_{\veck})$
at zero temperature. The straight line in (a) and (b) shows $E_{\veck}^{5}$
and $E_{\veck}^{3}$ dependence, repectively. The unit of $E_{\veck}$ is $D$,
half the band-width of conduction electrons.
\item[Fig.~13 \ \ ]
The resistivity $\rho_{\|}$ as a function of the temperature $T$,
in the unit $D$, half the band-width of conduction electrons. The parameters
in eq.(\ref{eqn:rho}) are chosen as $\rho_{0}=400$, $c_{1}=0.1$, $c_{2}=10$
and $c_{3}=5$. Closed circles show the temperature dependence of $\rho_{a}$
of the best sample of CeNiSn.~\cite{rf:Takabatake2,rf:Nakamoto}
\end{description}
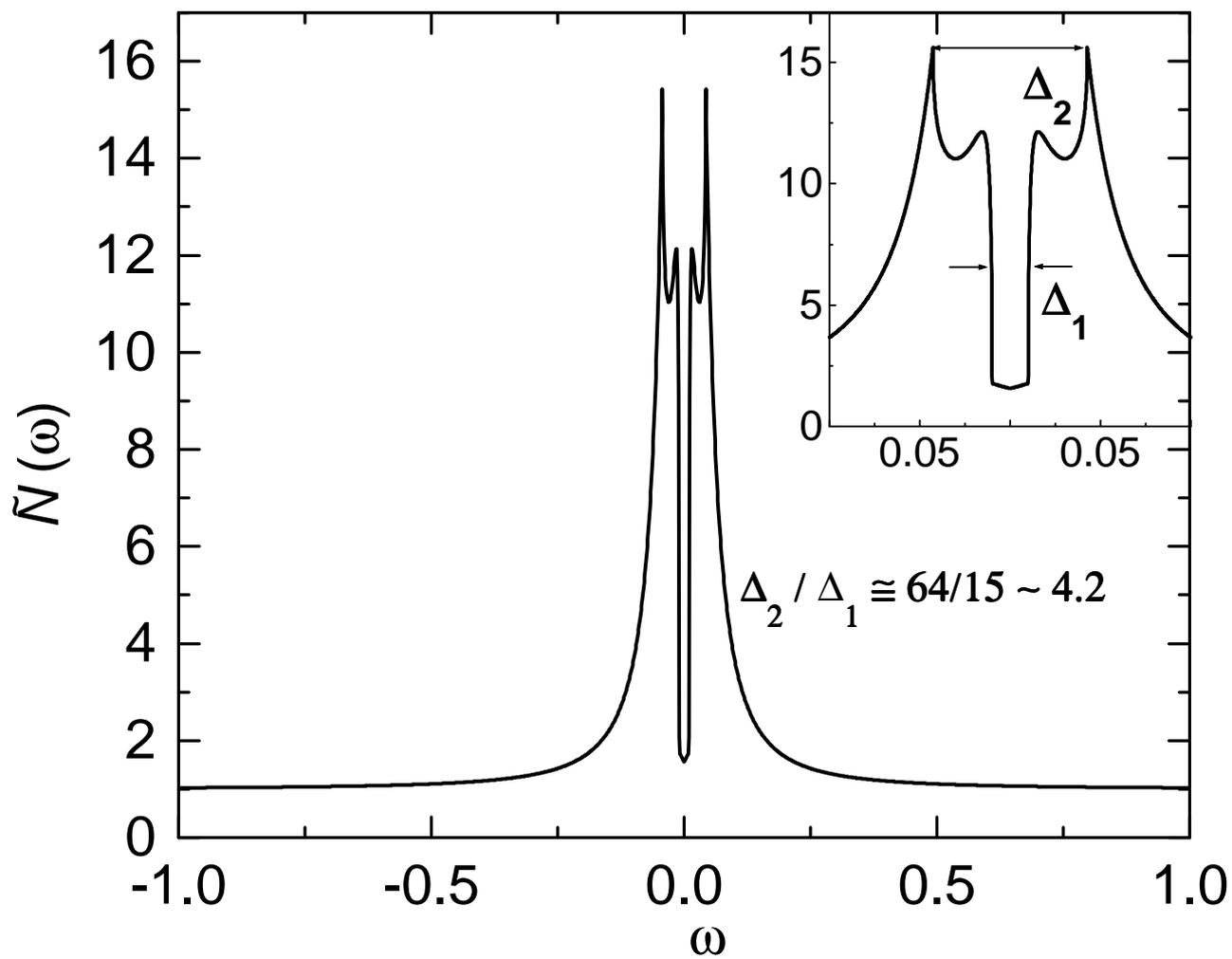
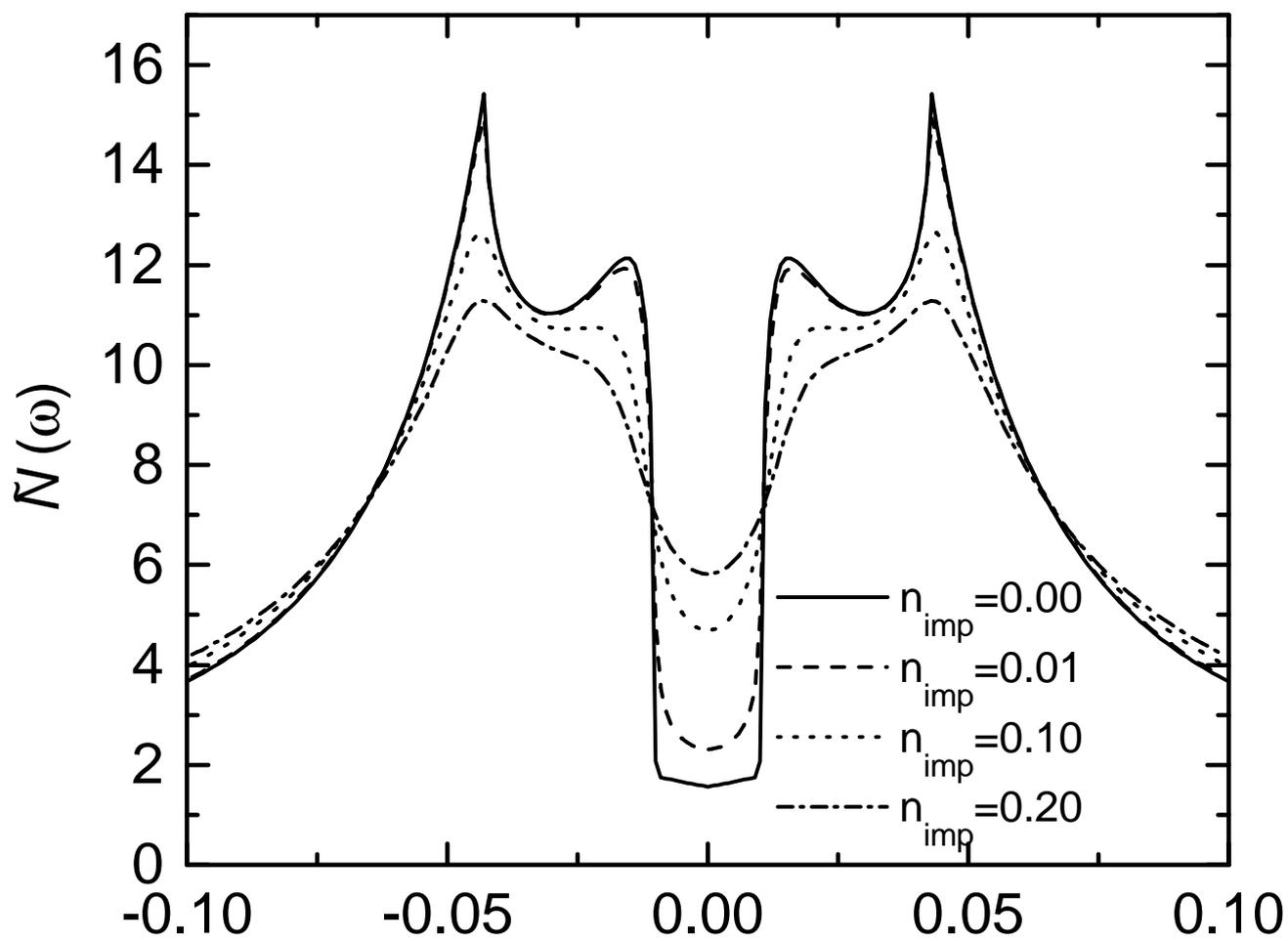
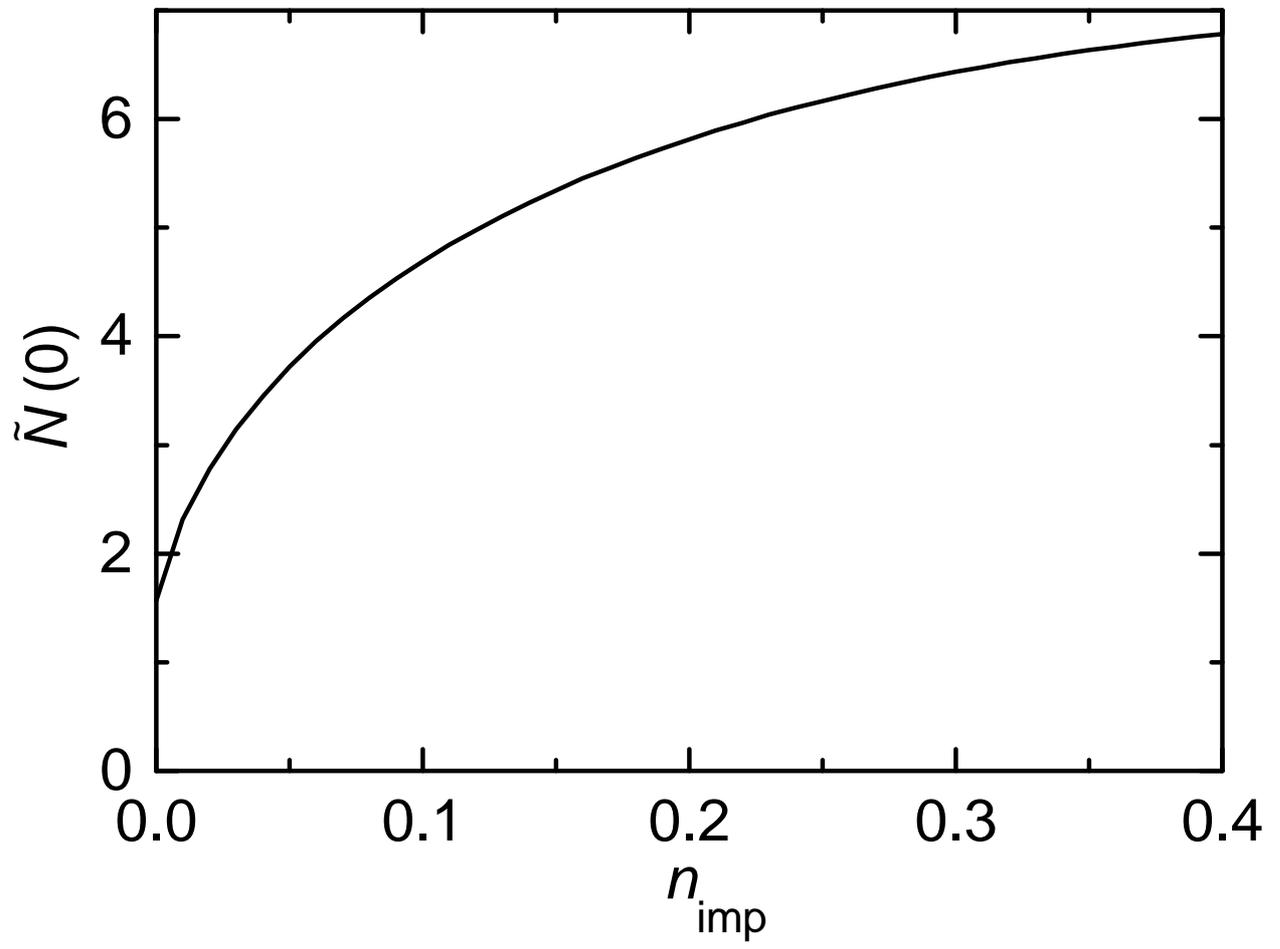
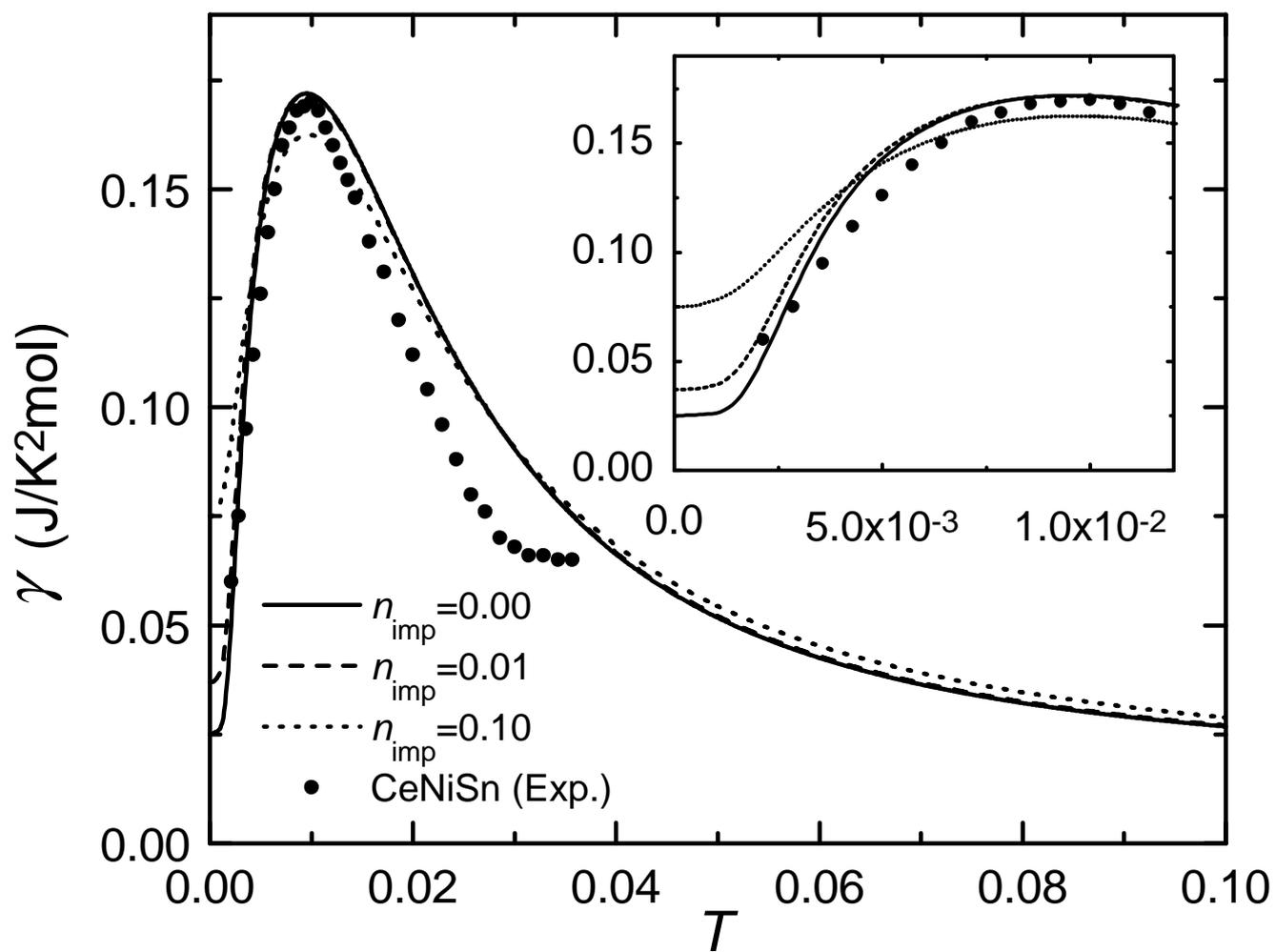
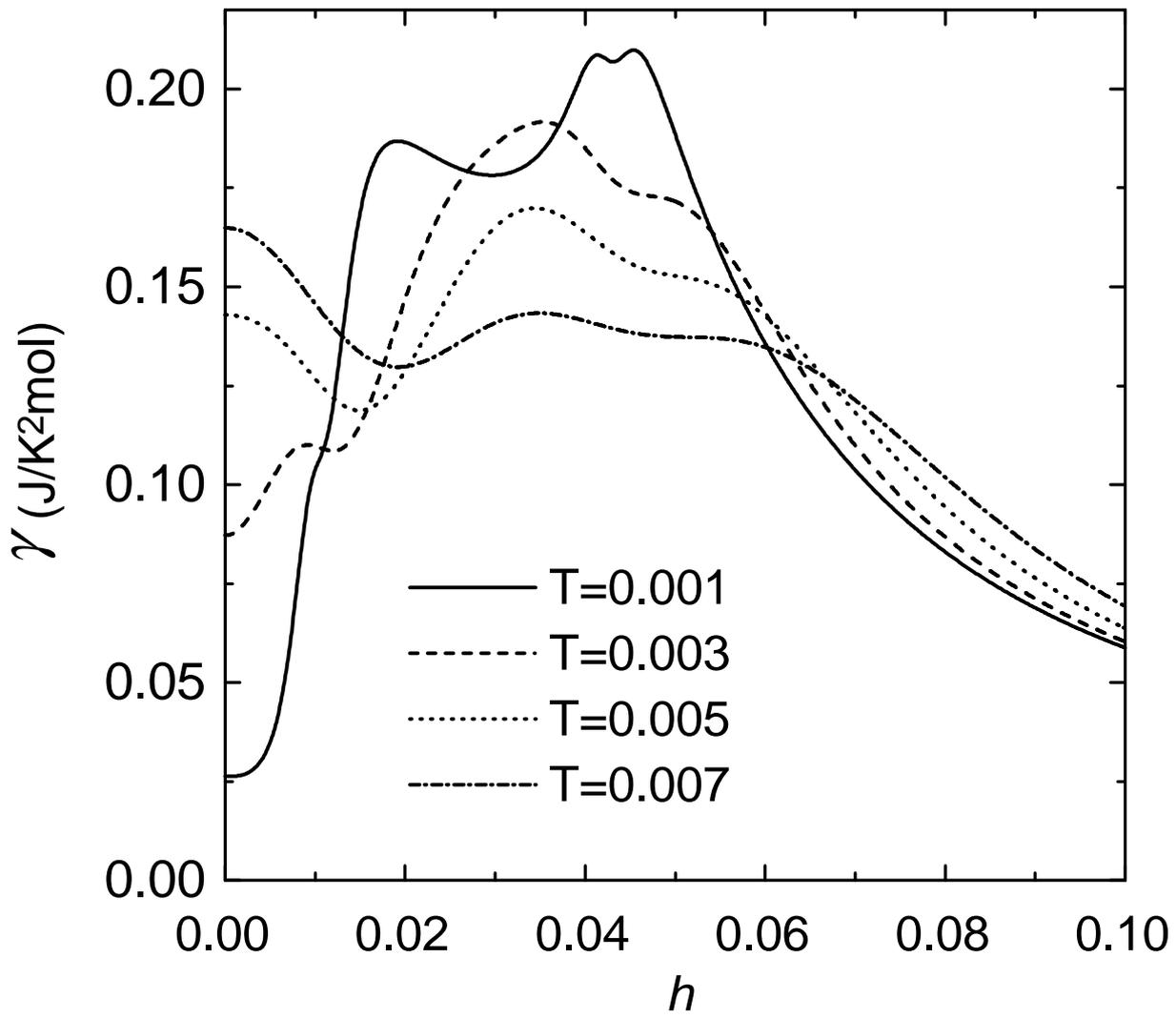
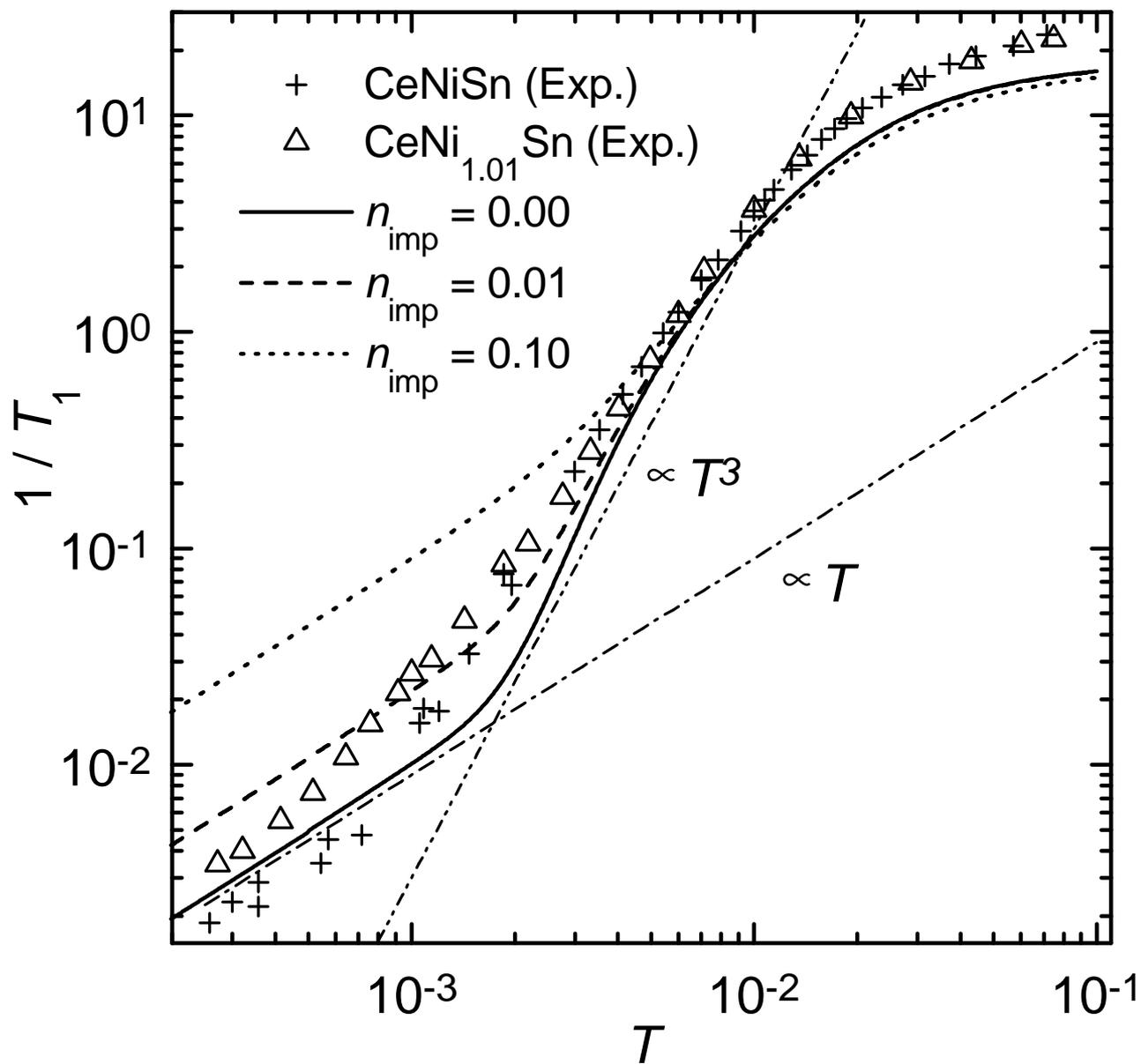
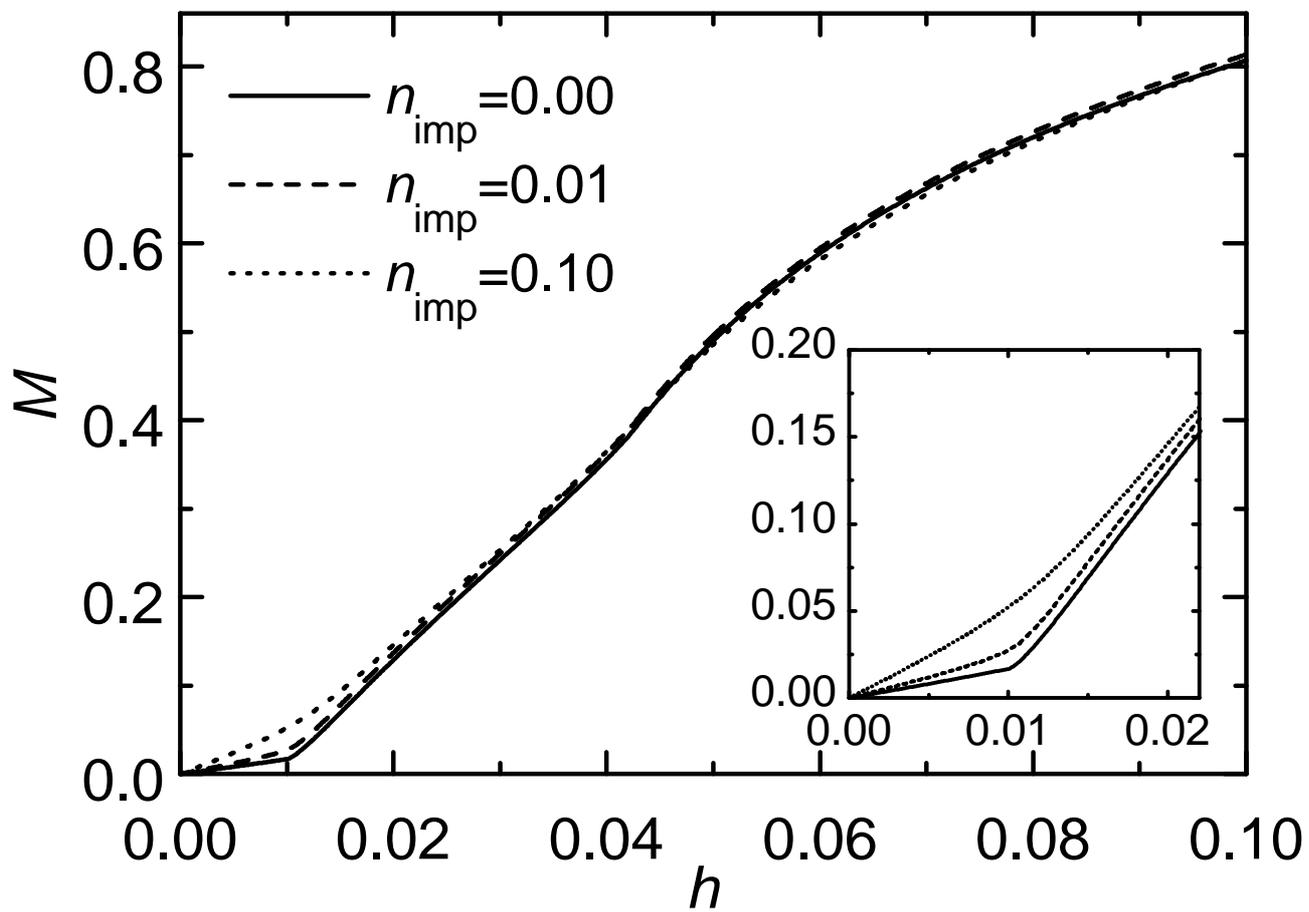
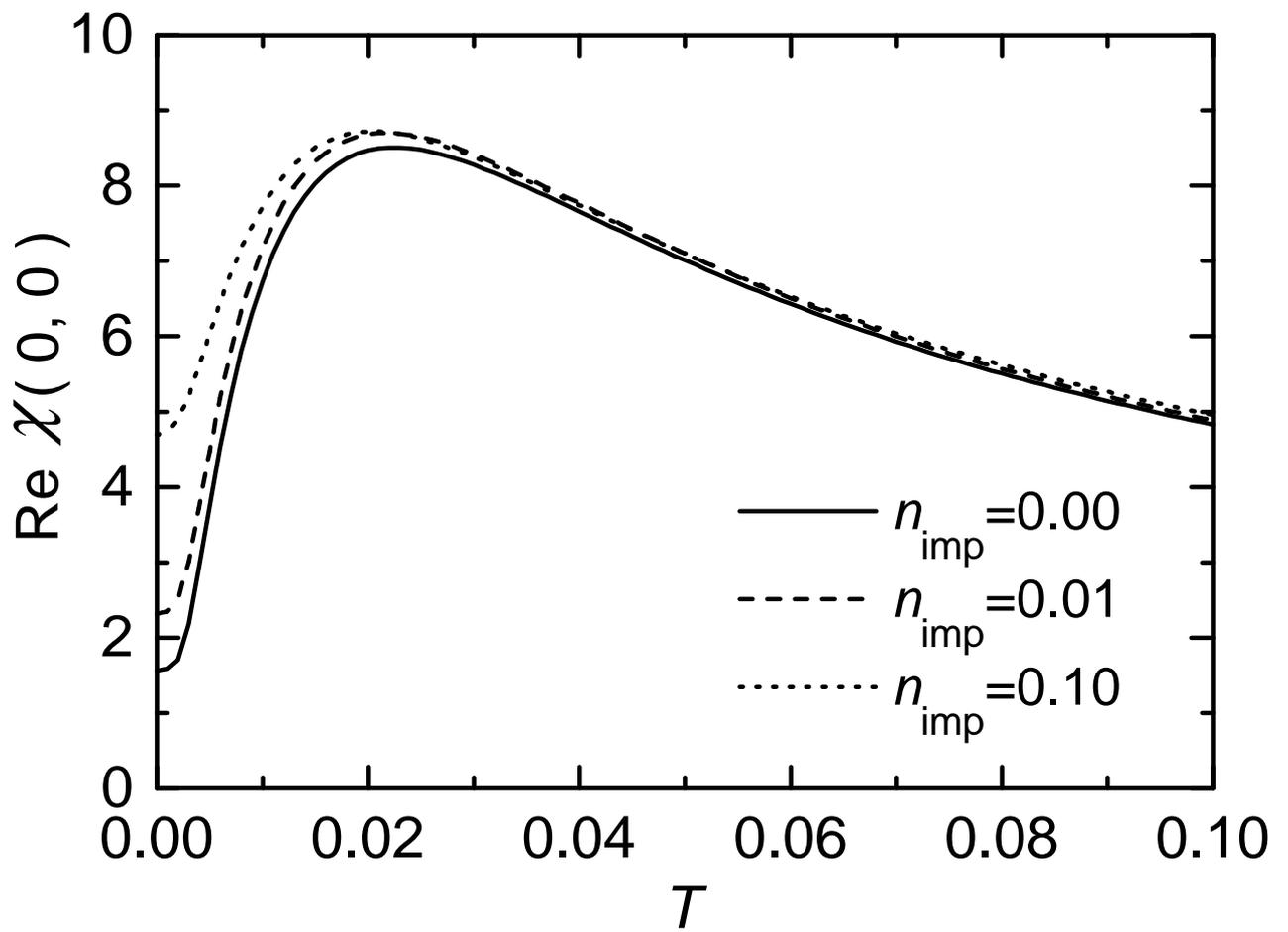
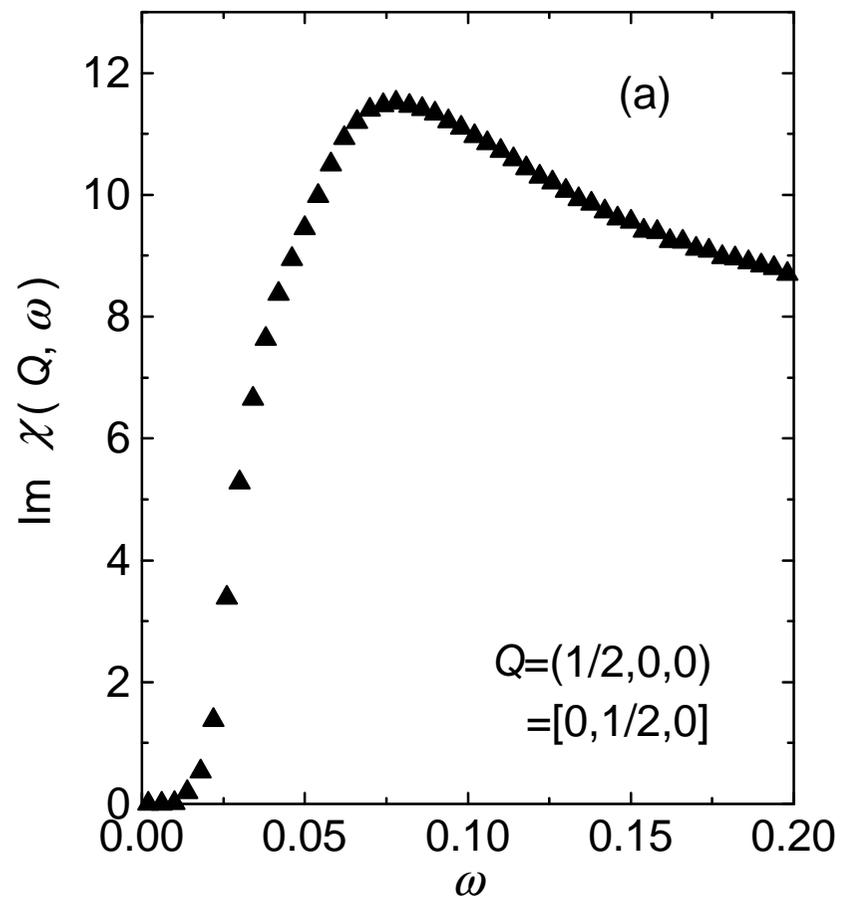
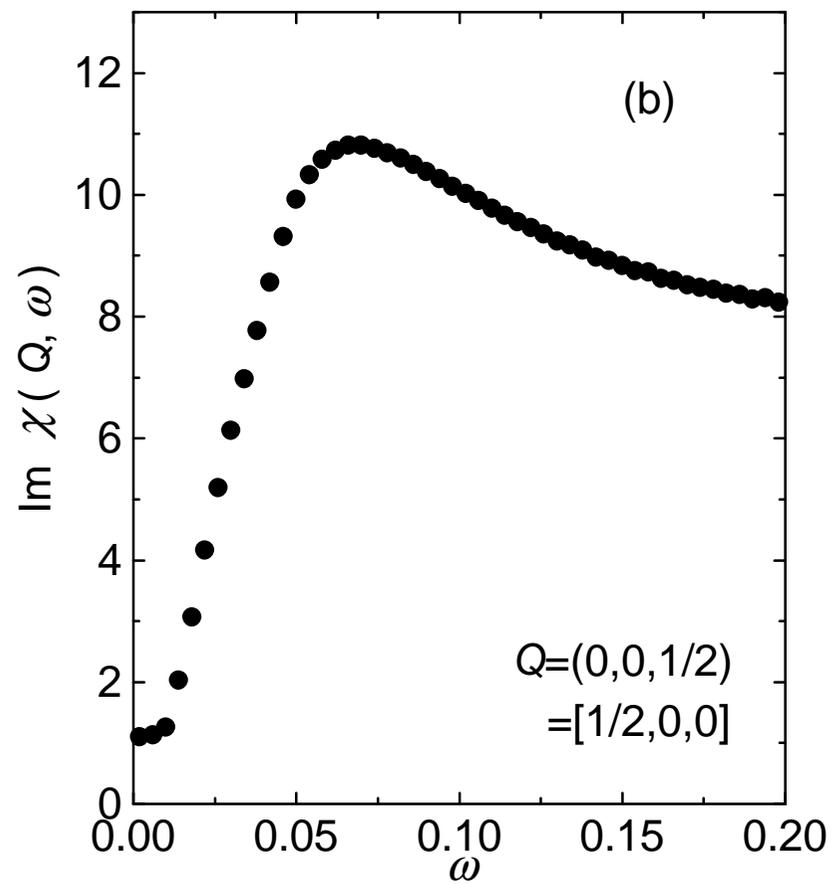
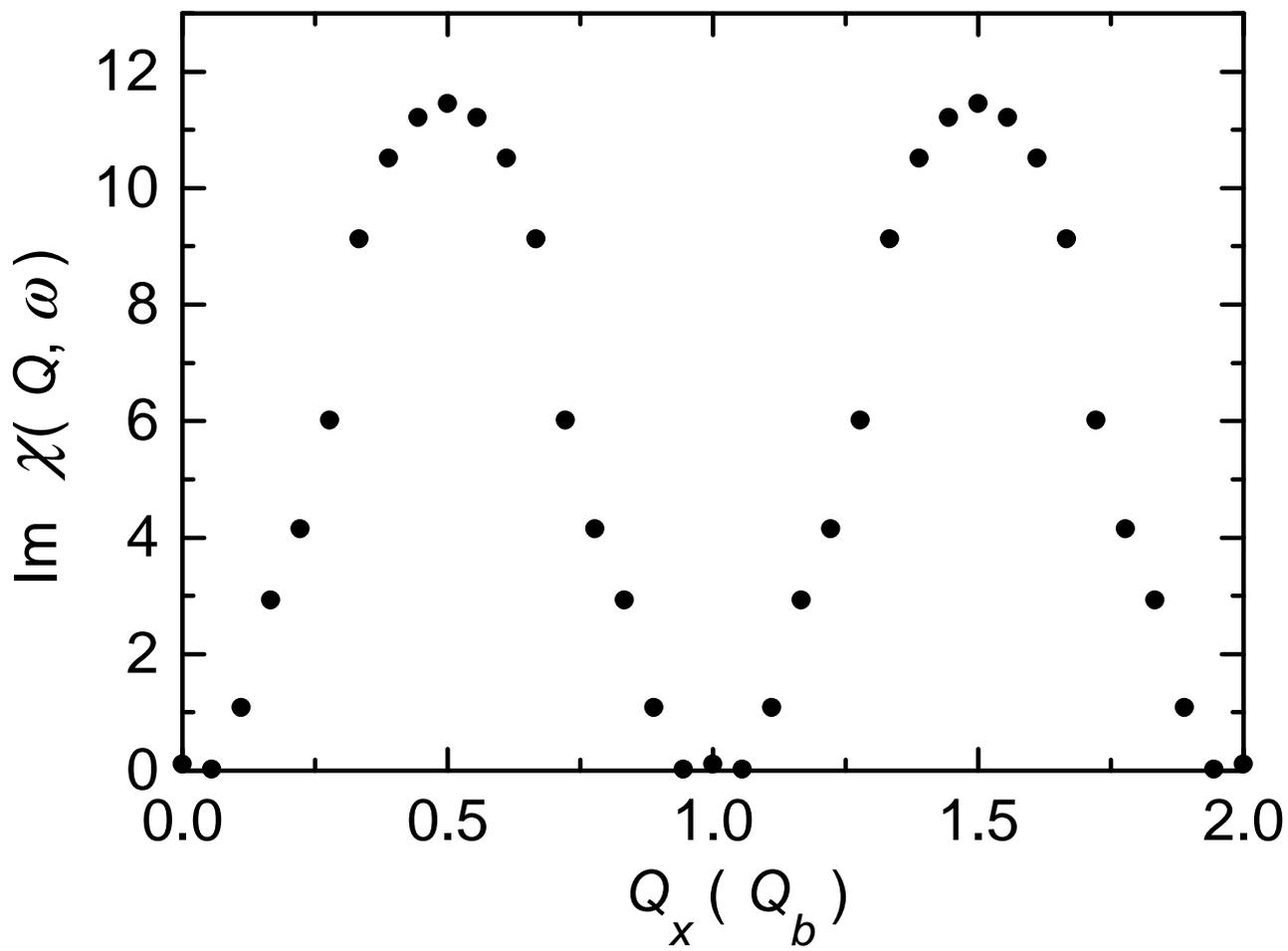
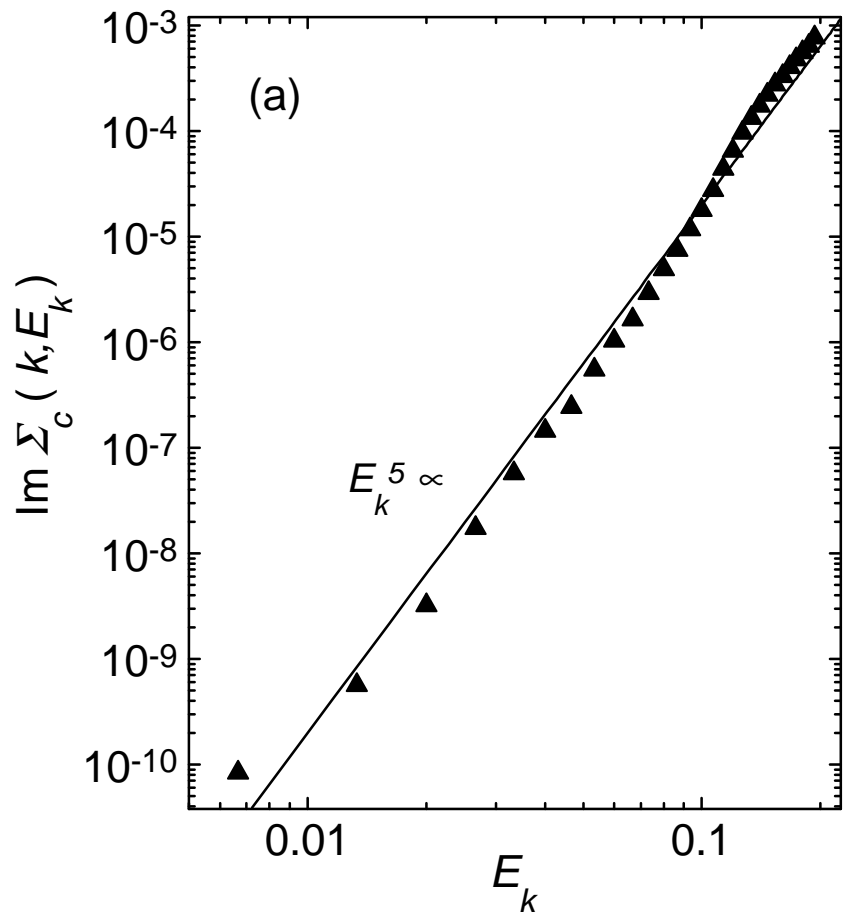
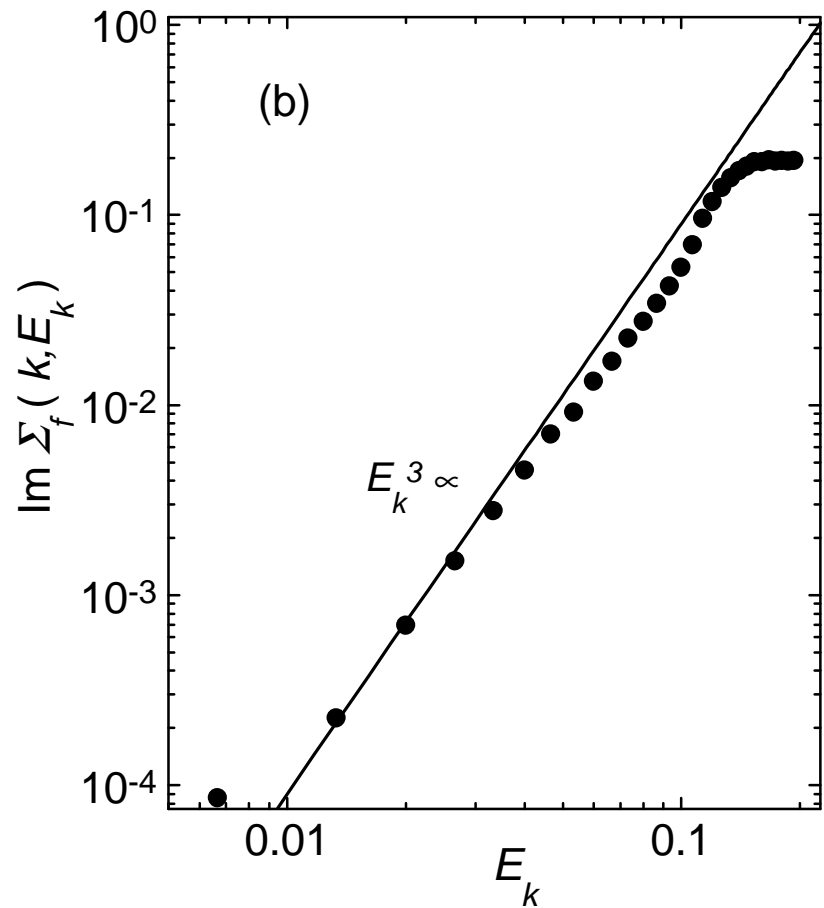
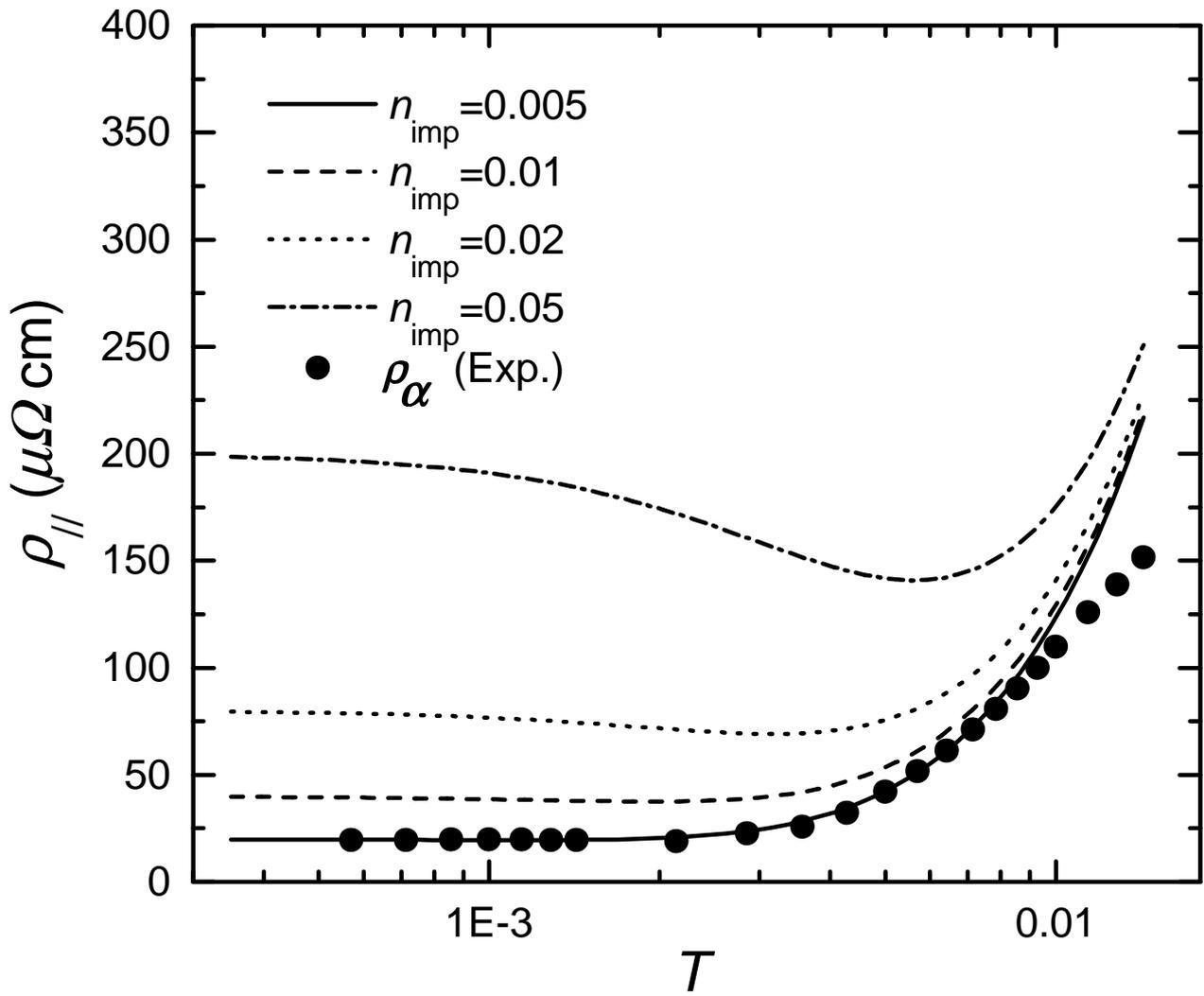
\end{document}